\documentclass[onepage,11pt,a4paper,pdftex]{article}

\usepackage{titletoc}
\usepackage{graphicx, amsfonts, amsmath, amssymb, amscd,  theorem, exscale,
epic, eepic, epsfig}

\usepackage{hyperref} 

\usepackage{makeidx}
\makeindex

\newcounter{Figure}
\theoremstyle{plain}
\theoremheaderfont{\scshape}

\newtheorem{Def}{\bf Definition}
\newtheorem{The}{\bf Theorem}

\theorembodyfont{\upshape}

\theoremheaderfont{\scshape\small} \theorembodyfont{\scshape\small} 
\newcommand{\real}{ {\mathbb R} }

\newcommand{\slap}{\mbox{$ \triangle \mkern -13mu / \ $}}
\newcommand{\nlap}{\mbox{$ \nabla \mkern -13mu / \ $}}
\newcommand{\dlap}{\mbox{$ div \mkern -13mu / \ $}}

\newcommand{\clap}{\mbox{$ curl \mkern -23mu / \ $}}

\newcommand{\philap}{\mbox{$ o \mkern -7mu / \mkern -10.8mu /  \ $}}

\newcommand{\be}{\begin{equation}}
\newcommand{\ee}{\end{equation}}
\newcommand{\bea}{\begin{eqnarray}}
\newcommand{\eea}{\end{eqnarray}}
\newcommand{\beas}{\begin{eqnarray*}}
\newcommand{\eeas}{\end{eqnarray*}}

\newcommand{\cDlap}{\mbox{$ \mathcal{D} \mkern -10mu / \ $}}
\newcommand{\Lie}{ {\mathcal L} }

\renewcommand{\S}{\mbox{$ \ ^{(S)}$}}

\begin{document}

\begin{center}
{\Large \bf An Extension of the Stability Theorem of the Minkowski 
\\ \vspace{5pt} 
Space 
in General Relativity} \\ 
\end{center}
\vspace{5pt}
\begin{center}
{\large \bf Lydia Bieri} \\ 
\end{center}
\begin{center}
{\large Department of Mathematics \\ 
Harvard University \\ 
Cambridge, MA 02138, USA. } \\ 
\end{center}
\begin{center}
{\large lbieri@math.harvard.edu} \\ 
\end{center}
\begin{center}
{\large December 30, 2008 } \\ 
\end{center}

\vspace{5pt}
\begin{center}
{\bf Abstract} \\ 
\end{center}
In this paper, we sketch the proof of the extension of the 
stability theorem of the Minkowski space in General Relativity done explicitly in \cite{lydia1}, \cite{lydia2}. 
We discuss solutions of the Einstein vacuum (EV) equations (obtained in the author's Ph.D. thesis \cite{lydia1} in 2007). 
We solve the Cauchy problem for more general, asymptotically flat initial data than 
in the pioneering work \cite{sta} of D. Christodoulou and S. Klainerman or than in any other 
work. 
Moreover, we describe precisely 
the asymptotic behaviour. 
Our relaxed assumptions on the initial data yield a spacetime 
curvature which is not bounded in $L^{\infty}(M)$. 
As a major result, we encounter in our work borderline cases, which we discuss in this paper as well. 
The fact that certain of our estimates are borderline in view of decay 
indicates that the conditions in our main theorem are sharp in so far as the assumptions on the decay  
at infinity on the initial data are concerned. 
Thus, the borderline cases are a consequence of our relaxed assumptions on the data, \cite{lydia1}, \cite{lydia2}. 
They are not present in the other works, as all of them place stronger assumptions on their data. 
We work with an invariant formulation of the EV equations. 
Our main 
proof is based on a bootstrap argument. 
To close the argument, we have to 
show that the spacetime curvature and the 
corresponding geometrical quantities have the required decay. 
In order to do so, the Einstein equations are decomposed with respect to specific foliations 
of the spacetime. 
This result generalizes the work \cite{sta} of D. Christodoulou and S. Klainerman.  \\ \\ 
\tableofcontents

\section{Introduction and Main Results}
%
%
%
%
The laws of General Relativity (GR) are the Einstein equations 
linking the curvature of the spacetime to its matter content. 
\be \label{equsGRgRT}
G_{\mu \nu} \  := \ R_{\mu \nu} \ - \ \frac{1}{2} \ g_{\mu \nu} \ R \ = \ 2 \ T_{\mu \nu}  \ , 
\ee
(rationalized units $4 \pi G = 1$), where, for $\mu, \nu = 0,1,2,3$,  
$G_{\mu \nu}$ is called the Einstein tensor,  
$R_{\mu \nu}$ is the Ricci curvature tensor,  
$R$ the scalar curvature tensor, 
$g$ the metric tensor and 
$T_{\mu \nu}$ denotes the energy-momentum tensor. \\ \\ 
This paper discusses the main results and steps of the proof of \cite{lydia1}, \cite{lydia2}, 
dealing with 
the global, nonlinear stability of solutions of the Einstein vacuum equations 
in General Relativity. 
The case of (\ref{equsGRgRT}) where $T_{\mu \nu} = 0$, are 
the Einstein vacuum (EV) equations. These read as follows: 
\be \label{Ricvac0}
R_{\mu \nu} \ = \ 0  \ . 
\ee 
Solutions of the EV equations 
are 
spacetimes $(M, g)$, where $M$ is a four-dimensional, oriented, differentiable manifold and $g$ 
is a Lorentzian metric obeying the EV equations. 
We study these equations for asymptotically flat systems. 
These are solutions where $M$ looks like flat Minkowski space with 
diagonal metric $\eta = (-1, +1, +1, +1)$ outside of spatially compact regions. 
Many physical cases require to study the Einstein equations in vacuum. 
Isolated gravitating systems such as binary stars, clusters of stars, galaxies etc. 
can be described in GR 
by asymptotically flat solutions of these equations. For, they can be thought of as having 
an asymptotically flat region outside the 
support of the matter. \\ \\ 
In view of the EV equations (\ref{Ricvac0}), it is an 
open problem, what is the sharp criteria 
for non-trivial asymptotically flat initial data sets to yield a maximal development that is complete. 
We generalize the 
results of D. Christodoulou and S. Klainerman in their joint work 
`The global nonlinear stability of the Minkowski space' \cite{sta}:  
Every strongly asymptotically flat, maximal, initial data which is globally close to the trivial data gives rise to 
a solution which is a complete spacetime tending to the Minkowski spacetime at infinity along any geodesic'. \\ \\ 
We solve the Cauchy problem with more general, asymptotically flat initial data. In particular, we have one 
less power of $r$ decay at spatial infinity and one less derivative than in \cite{sta}. We prove that also in this case, 
the initial data, under appropriate smallness conditions,  
yields a solution which is a complete spacetime, tending to the Minkowksi spacetime at 
infinity along any geodesic. 
In accordance with the initial data, the asymptotic flatness is 
correspondingly weaker. 
Contrary to the situation in \cite{sta}, 
certain estimates in our proof are borderline in view of decay, indicating that the 
conditions in our main theorem on the decay at infinity on the initial data are sharp. \\ \\ 
Our main results are stated 
in the Theorems \ref{maintheoremlb2} 
and \ref{maintheoremlb*1}.  \\ \\ 
We construct global solutions $(M, g)$ of the EV equations (\ref{Ricvac0}) 
for initial data specified in definition \ref{intAFB} below.  We use two foliations given by a 
maximal time function $t$ and an optical function $u$, respectively. The time function $t$ 
foliates our 4-dimensional spacetime into 3-dimensional spacelike hypersurfaces $H_t$, being 
complete Riemannian manifolds. Whereas the optical function $u$ induces a foliation 
of $(M,g)$ into null hypersurfaces $C_u$, which we shall refer to as null cones. 
The intersections $H_t \cap C_u = S_{t,u}$ are 2-dimensional compact Riemannian manifolds.  \\ 
\begin{Def}
An initial data set is a triplet $(H, \bar{g}, k)$ with $(H, \bar{g})$ being 
a three-dimensional complete Riemannian manifold and $k$ a two-covariant symmetric tensorfield on 
$H$, satisfying the constraint equations: 
\beas
\nabla^i \ k_{ij} \ - \ \nabla_j \ tr k \ & = & \ 0 \\ 
\bar{R} \ - \ \mid k \mid^2 \ + \ (tr k)^2 \ & = & \ 0 \ .
\eeas
\end{Def}
The constraint equations constrain the initial data. 
We recall that a development of an initial data set is an EV spacetime $(M, g)$ together with an 
imbedding 
$i : H \to M$ such that $g$ and $k$ are the induced first and second fundamental forms of $H$ in $M$. 
The barred quantities denote the metric and curvatures on $H$.  \\ \\ 
We work with a maximal time function $t$. 
That is, the level sets $H_t$ of the time function $t$ are required to be 
maximal spacelike hypersurfaces. Thus, they fulfill the equation 
$tr k = 0$. 
(See below.) 
Also, the lapse function $\Phi$ is introduced after the definition \ref{intdeftimet} of a time function. 
For a time function $t$ (that is, $dt \cdot X > 0$ for all future-directed timelike vectors $X$ at all points $p \in M$) 
the corresponding lapse function $\Phi$ is given by 
$ \Phi \ : = \ (-g^{\mu \nu } \partial_{\mu} t \partial_{\nu} t )^{- \frac{1}{2}}$. 
From the structure equations with respect to the $t$-foliation, using the EV equations (\ref{Ricvac0}), 
we derive 
the constraint equations, 
the evolution equations, 
and lapse equation below. 
The structure equations consisting of the variation, Codazzi and the trace of the Gauss equations, 
relate the spacetime curvature $R_{\alpha \beta \gamma \delta}$ to 
the Ricci curvature $\bar{R}_{ij}$ of $H_t$, the second fundamental form $k$ and the 
lapse function $\Phi$. 
Note that in the $3$-dimensional leaf $H_t$, its Ricci curvature $\bar{R}_{ij}$ completely determines 
the induced Riemannian curvature tensor $\bar{R}_{ijkl}$ as follows 
($\bar{R}$ is the scalar curvature $\bar{g}^{ij} \bar{R}_{ij}$):  
\be
\bar{R}_{ijkl} \ = \ 
\bar{g}_{ik} \ \bar{R}_{jl} \ + \ \bar{g}_{jl} \ \bar{R}_{ik} \ - \ \bar{g}_{il} \ \bar{R}_{jk} \ - \ 
\frac{1}{2} \ (\bar{g}_{ik} \ \bar{g}_{jl} \ - \ \bar{g}_{jk} \ \bar{g}_{il}) \ \bar{R}  \ . 
\ee
At this point, let us give the following formulas for the frame field 
$(e_0, e_1, e_2, e_3)$ for $M$, where 
$e_0 = \frac{1}{\Phi} T$ denotes the future-directed unit normal to the $H_t$ and 
$(e_1, e_2, e_3)$ is an orthonormal frame 
tangent to the leaves of the foliation: 
\bea
D_i e_0 \ & = & \  k_{ij} \ e_j \\ 
D_i e_j \ & = & \ \nabla_i e_j \ + \ k_{ij}\  e_0 \\ 
D_0 e_0 \ & = & \ (\Phi^{-1} \nabla_i \Phi) \ e_i \\ 
D_0 e_i \ & = & \ \bar{D}_0 e_i \ + \ (\Phi^{-1} \nabla_i \Phi) \ e_0  \ , 
\eea
$\bar{D}_0 e_i$ denoting the projection of $D_0 e_i$ to the tangent space of the foliation. 
Note that in a so-called Fermi propagated frame, it is 
$\bar{D}_0 e_i = 0$.\\ \\ 
Here, note that 
$g_{00} = -1$, $g_{0i} = 0$ and $g_{ij} = \bar{g}_{ij} = g(e_i, e_j)$ for 
$i, j = 1,2,3$. \\ \\ 
With respect to the foliation of the spacetime by a maximal time function $t$, 
the {\itshape constraint equations} take the following form: 
\bea
tr k \ & = & \ 0   \label{maxtrk0} \\ 
\nabla^i  k_{ij}  \ &  = & \ 0  \label{maxconstr1} \\ 
\bar{R}  \ & = & \  |k|^2  \ \ .  \label{maxconstr2}
\eea
The {\itshape evolution equations} for a maximal foliation are: 
\bea
\frac{\partial \bar{g}_{ij}}{\partial t} \ & = & \ 2 \Phi k_{ij}   \label{maxevol1} \\ 
\frac{\partial k_{ij}}{\partial t} \ & = & \  \nabla_i  \nabla_j \Phi  
\ - \ (\bar{R}_{ij} \  - \ 2 k_{im} k^m_j) \ \Phi \ \ .  \label{maxevol2}
\eea
Moreover, the 
{\itshape lapse equation} reads: 
\bea
\triangle \ \Phi \ & = & \ \mid k \mid^2 \ \Phi \ \ .   \label{maxlapse}
\eea
In our work, we consider asymptotically flat initial data of the following form: 
\begin{Def} \label{intAFB} 
(AFB) 
We define an asymptotically flat initial data set to be a 
AFB initial data set, if it is 
an asymptotically flat initial data set $(H_0, \bar{g}, k)$, 
where $\bar{g}$ and $k$ are sufficiently smooth 
and 
for which there exists a coordinate system $(x^1, x^2, x^3)$ in a neighbourhood of infinity such that with 
$r = (\sum_{i=1}^{3} (x^i)^2 )^{\frac{1}{2}} \to \infty$, it is:  
\bea
\bar{g}_{ij} \ & = & \ \delta_{ij} \ + \ 
o_3 \ (r^{- \frac{1}{2}}) \label{afgeng}  \\
k_{ij} \ & = & \ o_2 \ (r^{- \frac{3}{2}})    \ .    \label{afgenk}  
\eea
\end{Def} 
The initial data $(H_0, \bar{g}, k)$ has to satisfy the global smallness assumption below. 
We introduce $Q(a, x_{(0)})$ for the terms that have to be controlled by a small positive $\epsilon$. 
At a later point in the proof, $\epsilon$ has to be taken suitably small, depending on other quantities. 
\bea
Q(a, x_{(0)}) \ & = & \ 
a^{-1} \ \big( \ 
\int_{H_0} \ \big( \ 
\mid k \mid^2 \ + \ (a^2 + d_0^2) \ \mid \nabla k \mid^2  \nonumber \\ 
& & \ \ \ \ \quad \quad \ \ + \ (a^2 + d_0^2)^2 \ \mid \nabla^2 k \mid^2 \ \big)  \ d \mu_{\bar{g}} \nonumber  \\ 
& & 
\ \ + \ 
\int_{H_0} \ \big( \ 
(a^2 + d_0^2) \ \mid Ric \mid^2   \nonumber  \\ 
& & \ \ \ \ \quad \quad \ \ + \ (a^2 + d_0^2)^2  \ \mid \nabla Ric \mid^2 \ \big)  \ d \mu_{\bar{g}} \ \big)  \ ,  
\label{globalsalbQ}
\eea
where $a$ is a positive scale factor, and $d_0$ denotes the distance function from an 
arbitrarily chosen origin $x_{(0)}$. \\ \\ 
Let $\inf_{x_{(0)}, a} Q (x_{(0)}, a)$ denote the infimum over all choices of origin $x_{(0)}$ and all 
$a$ of the quantity defined by (\ref{globalsalbQ}). \\ \\ 
We consider asymptotically flat initial data sets for which the metric $\bar{g}$ 
is complete and there exists 
a small positive $\epsilon$ 
such that 
\be
\inf_{x_{(0)}, a} Q (x_{(0)}, a)  \  <  \ \epsilon \ .  \label{globalsalb2}  
\ee
One version of our main theorem is the following: 
\begin{The} \label{maintheoremlb2}
Any asymptotically flat, maximal initial data set, with complete metric $\bar{g}$, 
satisfying inequality (\ref{globalsalb2}), where the $\epsilon$ has to be taken sufficiently small, 
leads to a unique, globally hyperbolic, smooth and geodesically complete solution of the EV equations, 
foliated by the level sets of a maximal time function. This development is globally asymptotically flat. 
\end{The}
For later reference, we state the global smallness assumption B as follows. \\ \\ 
{\bf Global Smallness Assumption B}: \\
An asymptotically flat initial data set satisfies the 
{\bf global smallness assumption B}, if the metric $\bar{g}$ is complete and there exists 
a sufficiently small positive $\epsilon$ 
such that 
\be
\inf_{x_{(0)}, a} Q (x_{(0)}, a) \  <  \ \epsilon \ .  \label{globalsalb1}  
\ee
The global smallness assumption B 
has to be considered together with the main theorem \ref{maintheoremlb*1} 
in section \ref{intsectstatementresult*1}. Then, $\epsilon$ in (\ref{globalsalb1}) 
has to be taken suitably small such that the inequalities stated in the main theorem \ref{maintheoremlb*1} hold. 
This main theorem \ref{maintheoremlb*1} is the most precise statement of our results. \\ \\ 
To prove this result (theorem \ref{maintheoremlb2}, respectively theorem \ref{maintheoremlb*1}), 
we do not need any preferred coordinate system, but we rely on the 
invariant formulation of the EV equations. Also, the 
\begin{itshape} asymptotic behaviour \end{itshape} is given in a 
\begin{itshape} precise \end{itshape} way. \\ \\ 
We remark that by 
{\itshape geodesically complete} is denoted what in GR is called 
{\itshape g-complete} which means that every causal geodesic can be extended for all parameter values. \\ \\ 
At this point, let us recall the result of D. Christodoulou and S. Klainerman \cite{sta}. 
They consider the following 
strongly asymptotically flat initial data set: 
\begin{Def} \label{intSAFCK} 
(SAFCK) 
We define a strongly asymptotically flat initial data set in the sense 
of \cite{sta} (studied by Christodoulou and Klainerman) and in the following denoted 
by SAFCK initial data set, to be 
an initial data set $(H, \bar{g}, k)$, where 
$\bar{g}$ and $k$ are sufficiently smooth and there exists a coordinate system 
$(x^1, x^2, x^3)$ defined in a neighbourhood of infinity such that, \\
as 
$r = (\sum_{i=1}^3 (x^i)^2 )^{\frac{1}{2}} \to \infty$, 
$\bar{g}_{ij}$ and $k_{ij}$ are: 
\bea
\bar{g}_{ij} \ & = & \ (1 \ + \ \frac{2M}{r}) \ \delta_{ij} \ + \ o_4 \ (r^{- \frac{3}{2}}) \label{safg} \\ 
k_{ij} \ & = & \  o_3 \ (r^{- \frac{5}{2}}) \ ,  \label{safk}
\eea
where $M$ denotes the mass. 
\end{Def}
In order to state their global smallness assumption, Christodoulou and Klainerman introduce 
a 
quantity $Q_{CK}(x_{(0)}, b)$ that has to be controlled by a small positive $\epsilon$. It is 
\bea
Q_{CK}(x_{(0)}, b) \ & = & \ 
\sup_H \ \big( \ b^{-2} \ (d_0^2 \ + \ b^2)^3 \ \mid Ric \mid^2  \ \big)  \nonumber   \\ 
\ & & \ + \ 
b^{-3} \ 
\big( \ \int_H \ \sum_{l=0}^3 \ (d_0^2 \ + \ b^2)^{l+1} \ \mid \nabla^l k \mid^2  \nonumber \\ 
\ & & \ \ \ \ + \ \int_H \   \sum_{l=0}^1 \   (d_0^2 \ + \ b^2)^{l+3} \ 
 \mid \nabla^l B \mid^2      \ \big)   \label{QforglobalsaCK1}
\eea
with  
$d_0(x) = d(x_{(0)}, x)$ being the Riemannian geodesic distance between the point $x$ and a given 
point $x_{(0)}$ on $H$. 
$b$ is a positive constant, 
$\nabla^l$ denotes the $l$-covariant derivatives, and 
$B$ (Bach tensor) is the following symmetric, traceless $2$-tensor
\[
B_{ij} \ = \ \epsilon_j^{\ ab} \ \nabla_a \ ( R_{ib} \ - \ \frac{1}{4} \ g_{ib} \ R ) \ . 
\]
It is used to formulate the following global smallness assumption in \cite{sta}. \\ \\ 
{\bf Global Smallness Assumption CK:} \\
A strongly asymptotically flat initial data set is said to satisfy the {\bf global smallness assumption CK} 
if the metric 
$\bar{g}$ is complete and there exists a sufficiently small positive $\epsilon$ such that 
\be \label{globalsmallnessassptck}
\inf_{x_{(0)} \in H, b \geq 0} \ Q_{CK}(x_{(0)}, b)  \ < \ \epsilon \ . 
\ee
Then, one version of the main theorem in \cite{sta}, 
'The global nonlinear stability of the Minkowski space', 
by Christodoulou and Klainerman is 
stated as follows: 
\begin{The} {\bf (D. Christodoulou and S. Klainerman, \cite{sta}, p. 17, Theorem 1.0.3)} \label{maintheoremCK2}
Any strongly asymptotically flat, maximal, initial data set that satisfies 
the global smallness assumption CK (\ref{globalsmallnessassptck}), leads to a unique, globally hyperbolic, 
smooth and geodesically complete solution of the EV equations foliated by a normal, maximal 
time foliation. 
This development is globally asymptotically flat. 
\end{The}
The full version of their result, Christodoulou and Klainerman provide in 
\cite{sta}, p.298, Theorem 10.2.1. \\ \\ 
There is no additional restriction on the data. 
The authors do not use a preferred coordinate system, but their 
proof relies on the invariant formulation of the EV equations. 
Moreover, they obtain a 
\begin{itshape}precise description \end{itshape} of the 
\begin{itshape}asymptotic behaviour at null infinity. \end{itshape}  \\ \\ 
Our initial data ((\ref{afgeng}), (\ref{afgenk}))  is more general than the one in ((\ref{safg}), (\ref{safk})) in the sense that 
in (\ref{afgeng}) and in (\ref{afgenk}) we have one less derivative and less fall-off 
by one power of $r$ than 
in ((\ref{safg}), (\ref{safk})). 
We show {\bf existence} and {\bf uniqueness} of solutions of the EV equations 
under these {\bf relaxed assumptions} on the 
initial data (AFB, see (\ref{globalsalb1})). 
As we are assuming less on our initial data, 
the description of the {\bf asymptotic behaviour} of the 
curvature components is less precise than  
in \cite{sta} (with (\ref{safg}), (\ref{safk})). 
However, it is as precise as it can be with these 
relaxed assumptions. 
The case we study does not tend as fast to Minkowski as the situation in \cite{sta}. 
In our proof, we use the main structure as in the proof of \cite{sta}, namely a bootstrap argument. 
However, the proof itself and the techniques differ considerably from the original one. 
Our more general case requires subtle and different treatment of the most delicate estimates. 
Another major difference to the situation studied in \cite{sta} by Christodoulou and Klainerman, 
and which arises from our relaxed assumptions, is the fact that we 
encounter {\bf borderline cases in view of decay} in the power of $r$, indicating that 
the conditions in our main theorem on the decay at infinity of the initial data are {\bf sharp}. 
Any further relaxation would make the corresponding integrals diverge and the argument would 
not close any more. 
As a consequence from imposing less conditions on our data, the 
{\bf spacetime curvature is not in $L^{\infty}(M)$}. 
We only control one derivative of the curvature (Ricci) in $L^2 (H)$. 
By the trace lemma, the Gauss curvature $K$ in the leaves of the $u$-foliation $S$ is only 
in $L^4(S)$. Contrary to that, in \cite{sta}, the Ricci curvature is in $L^{\infty} (H)$, 
and in $L^{\infty}(S)$. The 
authors control two derivatives of the curvature (Ricci) in $L^2 (H)$. 
Thus, this is a disadvantage and an advantage. First, as we do not have the curvature 
bounded in $L^{\infty}$, certain steps of the proof become more subtle. On the other hand, 
we do not have to control the second derivatives of the curvature, which simplifies the proof considerably. 
A major simplification is the fact, that we do 
{\bf not use any rotational vectorfields} in our 
proof. We gain control on the {\bf angular derivatives} of the {\bf curvature directly from the Bianchi equations}. 
Whereas 
in \cite{sta}, a difficult construction of rotational vectorfields was necessary. 
Moreover, in our situation, {\bf energy} and {\bf linear momentum} are {\bf well-defined and conserved}, 
whereas the {\bf (ADM) angular momentum} is {\bf not defined}. 
This is different to the situation investigated in \cite{sta}, where 
all these quantities are well-defined and conserved.  \\ \\ 
The results of \cite{sta} 
yield the laws of gravitational radiation proposed by Bondi \cite{bon}.  
In particular, they explain 
the physical theory of the so-called memory effect (see \cite{chrmemory}) in the 
framework of gravitational radiation. 
D. Christodoulou discusses this in 
his paper \cite{chrmemory} about nonlinear nature of gravitation and gravitational-wave experiments. 
This memory effect is due to the nonlinear character of the asymptotic laws at 
future null infinity. 
The many well-known experiments to detect gravitational waves, going on and planned for 
the near future, build on this effect. 
In the same paper the formula for the power radiated to infinity at a given 
retarded time, in a given direction, per unit solid angle, is stated; as well as 
the formula for the total energy radiated to infinity 
in a given direction, per unit solid angle. \\ \\ 
The full version of our main theorem is stated in  
theorem \ref{maintheoremlb*1} in section \ref{intsectstatementresult*1}. 
Crucial steps of the proof of the main theorem, in particular the bootstrap argument, 
are given in 
section \ref{subchcrucialstepsproofBA***12}. 
In section \ref{proofoutline*2} we discuss the proof further and we investigate a borderline case. \\ \\ 
For the reader who wishes to delve deeper into the area or the background material on which developments 
in this area depend, then among the vast interesting literature we suggest the following in addition to the references cited in this paper: 
\cite{taubin1}, \cite{chrgls}, \cite{chrap},  \cite{stasurv}, \cite{cdgworld},  \cite{chrdglobalivps}, \cite{chrd2007shocks3d},  \cite{ChrKllinfe}, 
\cite{boost},  \cite{hawk}, \cite{heb1},  \cite{heb2}, \cite{hoer}, \cite{leeparker1}, \cite{newto},  
\cite{oss}, \cite{parktaub},  \cite{pen}, \cite{wccrp},  \cite{sccrp}, \cite{mistr1},  \cite{schyau}, \cite{witten}. \\ 
\section{Setting}
\label{chintsetting***s12}
The spacetime manifold $(M,g)$ is defined above. 
For a Lorentzian metric $g$, there exists a vector 
$V$ in $T_pM$ such that $g_p(V,V) < 0$. Its 
$g_p$-orthogonal complement is defined as 
$\Sigma_V = \{ X: \ g_p(X,V) = 0 \}$ and 
$g_p$ restricted to $\Sigma_V$ is positive definite. \\ \\ 
Then, at each $p$ in $M$ we can choose 
a positive orthonormal frame $(e_0, e_1, e_2, e_3)_p$ continuously. 
We obtain the positive orthonormal frame field 
consisting of $e_0, e_1, e_2, e_3$ with: 
\be
e_0 = \frac{V}{\sqrt{-g(V,V)}}
\ee
and $e_1, e_2, e_3$ 
being an orthonormal basis for $\Sigma_V$. \\ \\ 
A given vector $X$ in $T_pM$ can be expanded as 
\begin{eqnarray*}
X & = &  X^0 e_0 + X^1 e_1 + X^2 e_2 + X^3 e_3 \\
                  & = & \sum_{i} X^{i} e_{i} \ \ \ \ \ \ \ \ \ \  ( i = 0, 1, 2, 3).
\end{eqnarray*}
Consequently, it is  
\begin{eqnarray*}
g(e_{i}, e_{j}) = \eta_{i j} &  =  & diag(-1, +1, +1, +1). \\ 
g(X,X) & = & - (X^0)^2 + (X^1)^2 + (X^2)^2 + (X^3)^2 \\
  & = &  \sum_{i j} \eta_{ij} X^{i} X^{j}
\end{eqnarray*}
At a point $p$ in $M$, we distinguish three types of vectors. 
Namely, null, timelike and spacelike vectors. The null vectors form a double cone at $p$, while 
the timelike vectors form an open set of two connected components, that is, the interior of this cone, 
and the spacelike vectors a connected open set being the exterior of the cone. 
They are defined as follows. 
\begin{Def}
The null cone \index{null cone} (or light cone) at $ p $ in $M$ is 
\[
N_p \ = \ \{ X \neq 0 \ \ \in \ T_pM : \ g_p(X,X) = 0  \}  \ . 
\]
The double cone consists of $N_p^+$ and $N_p^-$: $N_p \ = \ N_p^+ \cup N_p^-$. 
\end{Def}
Denote by $I_p^+$ the interior of $N_p^+$ and by $I_p^-$ the interior of $N_p^-$. 
\begin{Def}
The set of timelike \index{timelike!} vectors at  $p$ in $M$ is given by 
\[
I_p \ := \  I_p^+ \ \cup \   I_p^- \  = \  \{ X \ \in \ T_pM : \ g_p(X,X) < 0 \}  \ . 
\]
\end{Def}
\begin{Def}
The set of spacelike \index{spacelike!} vectors at  $p$ in $M$ is defined to be  
\[
S_p \ := \   \{ X \ \in \ T_pM : \ g_p(X,X) > 0 \}  \ . 
\]
\end{Def}
Thus, $S_p$ is the exterior of $N_p$.  
\begin{Def} \label{intcausalcurve*1}
A causal curve \index{causal! causal curve} in $M$ is a differentiable curve $\gamma$ whose 
tangent vector $\dot{\gamma}$ at
each point $p$ in $M$ belongs to  $ I_p \cup N_p$, i.e. is either timelike or null. 
\end{Def}
\begin{Def}
The causal future of a point $p$ in $M$, denoted by $J^+(p)$, 
is the set of all points $q  \in  M$ for which there exists a 
future-directed causal curve initiating at $p$ and ending at $q$. 
\end{Def}
Correspondingly, we can define $J^-(p)$, the causal past of $p$. 
We also need the causal future of a set $S$ in $M$: 
\begin{Def}
The causal future $J^+(S)$ of any set $S \ \subset \ M$, 
\index{causal! future} \index{causal! past}
in particular in the case that $S$ is a closed set, is 
\[
J^+(S) \ = \ \{ q \ \in \ M : \ q \ \in \ J^+(p) \  \mbox{for some $p \ \in \ S$} \} \ . 
\] 
\end{Def}
Similarly, the definition is given for $J^-(S)$. 
The boundaries $\partial J^+(S)$ and $\partial J^-(S)$ of $J^+(S)$ and $J^-(S)$, respectively, 
for closed sets $S$ are null hypersurfaces. They are generated by null geodesic segments. 
The null geodesics \index{geodesics! null} 
generating $J^+(S)$ have past end points only on $S$. 
These null hypersurfaces $\partial J^+(S)$ and $\partial J^-(S)$ 
are realized as level sets of functions $u$ satisfying
the Eikonal equation $g^{\mu \nu} \partial_{\mu} u \partial_{\nu} u \ = \ 0$. 
\begin{Def}
$H$ is called a null hypersurface \index{null hypersurface} if at each point 
$x$ in $H$ the induced metric 
$g_x \mid T_xH$ is degenerate. 
\end{Def}
This means that there exists a $L \neq 0 \ \in \ T_xH$ such that 
\[
g_x(L, X) \ = \ 0 \ \ \ \ \forall  \ X \ \in \ T_xH \ . 
\]
Now, let $u$ be a function for which each of its level sets is a null hypersurface. 
Then we can set 
\[
L^{\mu} \ = \ - g^{\mu \nu} \partial_{\nu} u 
\]
and we have 
\[
g(L,L) \ = \ 0 .  
\]
The same reads in terms of $du$ as follows: 
$g^{\mu \nu} \partial_{\mu}u \partial_{\nu}u \ = \ 0$,  
which is the {\itshape Eikonal equation.} 
In fact, $L$ is a geodesic vectorfield, that is, the integral
curves of $L$ are null geodesics. 
A null hypersurface is generated by null geodesic segments. \\ \\ 
\begin{Def}
A hypersurface $H$ is called spacelike if at each $x$ in $H$, the induced metric 
\[
g_x \mid_{T_xH} \ =: \ \bar{g}_x 
\]
is positive definite. 
\end{Def}
We observe that,  
$(H, \bar{g})$ is a proper Riemannian manifold. 
And the $g$-orthogonal complement 
of $T_xH$ is a $1$-dimensional subspace of $T_xM$ on which $g_x$ is negative definite. 
Thus there exists a vector $N_x \ \in \ I_x^+$ of unit magnitude 
\[
g_x(N_x, N_x) \ = \ -1
\]
whose $span$ is this $1$-dimensional subspace. We refer to $N$ (the so-defined vectorfield along $H$) as the
future-directed unit normal to $H$. \\ \\ 
The second fundamental form of the hypersurface $H$ is denoted as 
\begin{equation}
k(X, Y) \ = \ g( D_X N, Y) \ \ \ \ \ \ \ \ \ \forall \ X, \ Y \ \in \ T_xH.
\end{equation}
A very important notion in GR and in this work is a Cauchy hypersurface, being defined with the help of causal curves: 
\begin{Def}
A Cauchy hypersurface \index{Cauchy! Cauchy hypersurface!} is a complete spacelike hypersurface $H$ in $M$ 
(i.e. $(H, \bar{g})$ is a 
complete Riemannian manifold) such that if $\gamma$ is any causal curve through any point 
$p \ \in \ M$, then $\gamma$ intersects $H$ at exactly one point.  
\end{Def}
A spacetime admitting a Cauchy hypersurface is called 
{\itshape globally hyperbolic.} \\ \\ 
Assuming the spacetime to be globally hyperbolic, a {\itshape time function} $t$ 
can be defined. 
\begin{Def} \label{intdeftimet}
Let the spacetime $(M,g)$ be globally hyperbolic. A time function is then a 
differentiable function $t$ such that 
\be
dt \ \cdot \ X \ > \ 0 
\ee
for all $X \ \in \ I_p^+$ and for all $ p \ \in \ M$. 
\end{Def}
The foliation given by the level surfaces $H_t$ of $t$ is called 
{\itshape $t$-foliation.} 
Denote by $T$ the following future-directed normal to this foliation: 
$
T^{\mu}  =  -  \Phi^2  g^{\mu \nu}  \partial_{\nu} t . 
$ 
It is $Tt = T^{\mu} \partial_{\mu} t = 1$. 
Now, a spacetime foliated in this fashion, 
is 
diffeomorphic to the product $\real \times \bar{M}$ where $\bar{M}$ is a 
$3$-manifold, each level set $H_t$ of $t$ being diffeomorphic to $\bar{M}$. 
The integral curves of $T$ are the orthogonal curves to the $H_t$-foliation. They are 
parametrized by $t$. 
Relative to this representation of $M$, the metric $g$ reads: 
\be
g \ = \ -  \Phi^2 dt^2 + \bar{g} 
\ee
with $\bar{g} = \bar{g}(t)$ denoting the induced metric on $H_t$. Note that 
$\bar{g}$ is positive definite. 
Here, $\Phi$ is the {\itshape lapse function} corresponding to the time function $t$. 
It is defined as follows: 
\be
\Phi \ : = \ (-g^{\mu \nu } \partial_{\mu} t \partial_{\nu} t )^{- \frac{1}{2}} \ . 
\ee
This lapse function measures the normal separation of the leaves $H_t$. 
By $N$ (already given above) denote the unit normal $N = \Phi^{-1}T$. Its integral curves are the same as for $T$, but 
parametrized by arc length $s$. \\ \\ 
In view of the first variational formula below, consider 
a frame field 
$e_1, e_2, e_3$ for $H_t$, Lie transported along the integral curves of $T$. That is, we have 
\[
[T, e_i] \ = \ 0 
\]
for $i = 1, 2, 3$. Denote $\bar{g}_{ij} = \bar{g} (e_i, e_j) = g(e_i, e_j)$. 
Then the {\itshape first variational formula} is: 
\begin{eqnarray}
k_{ij} \ & = & \ k(e_i, e_j) \\ 
         & = & \ \frac{1}{2 \Phi} \frac{\partial \bar{g}_{ij}}{\partial t} \ . 
\end{eqnarray}
One can choose a time function $t$, the level sets $H_t$ of which are 
{\itshape maximal} spacelike hypersurfaces. This eliminates the indeterminacy of the 
evolution equations. The definition \ref{intdeftimet} of a time function implies a freedom of choice. In fact, 
$t$ being subject only to 
$dt \ \cdot \ X \ > \ 0$ 
for all $X \ \in \ I_p^+$ and for all $ p \ \in \ M$, is arbitrary. 
We now fix our time function $t$ by the condition to be 
{\itshape maximal}. This means, we require the level sets $H_t$ of the time function $t$ to be 
{\itshape maximal spacelike hypersurfaces}. 
It describes the fact that any compact perturbation of $H_t$ decreases the volume. 
Thus, $H_t$ satisfies the maximal hypersurface equation 
\be
tr \ k \ \ = \ \ 0 \ \ . 
\ee
The existence of maximal surfaces in asymptotically flat spacetimes under 
slightly more general conditions, but for data with the same fall-off as ours has been 
proven by R. Bartnik, P.T. Chru\'sciel and N. O'Murchadha in \cite{bart3}. 
It was first proven by R. Bartnik for stronger fall-off in \cite{bart}.  \\ \\
\begin{Def}
A maximal time function is a time function $t$ whose level sets are 
maximal spacelike hypersurfaces, 
being complete and tending to parallel spacelike coordinate hyperplanes 
at spatial infinity. We also require that the associated lapse function $\Phi$ 
tends to $1$ at spatial infinity. 
\end{Def}
There is one such function up to an additive constant for each choice of family of 
parallel spacelike hyperplanes in Minkowski spacetime. 
These families are connected by the action of elements of the Lorentz group. \\ \\ 
One can single out one family by choosing 
\be
P^i \ = \ 0 \ \ .
\ee
Then the time function $t$ is unique up to an additive constant. \\ \\ 
The covariant differentiation on the spacetime $M$ is denoted by $D$. 
For that on $H$ write $\nabla$. 
Whenever a different notation is used, it is indicated. 
In the sequel, denote by $R$ the Riemannian curvature tensor of $M$, and by 
$\bar{R}$ the one of $H$. 
We shall work with the Weyl tensor $W$, not directly with the Riemannian curvature, 
for reasons explained below. \\ \\ 
As motivated at the beginning, we are studying asymptotically flat solutions of the EV equations 
(\ref{Ricvac0}): 
\[
R_{\mu \nu} \ = \ 0 \ . 
\]
Therefore, let us now explain what in general 
an asymptotically flat initial data set is. In view of the different types of asymptotic flatness, 
we first give the general definitions. Then, we can compare them with the definitions 
\ref{intAFB} and \ref{intSAFCK} from above describing the situations in \cite{sta} and 
\cite{lydia1}, respectively. \\ 

%
%
\begin{Def} \label{intasflat1}
A general asymptotically flat initial data set 
$(H, \bar{g}, k)$ is an  
initial data set such that 
\begin{itemize}
\item the complement of a compact set in 
$H$ is diffeomorphic to the complement of a closed ball in $\real^3$ 
\item and there exists 
a coordinate system $(x^1, x^2, x^3)$ 
in this complement relative to which the metric 
components 
\beas
\bar{g}_{ij} \ & \ \rightarrow \ & \ \delta_{ij} \\ 
k_{ij} \ & \ \rightarrow \ & \  0  
\eeas
sufficiently 
rapidly 
as 
$r = (\sum_{i=1}^3 (x^i)^2 )^{\frac{1}{2}} \to \infty$. 
\end{itemize}
\end{Def}
Generally, one defines 
'strong' asymptotic flatness as follows: 
\index{asymptotically! strongly asymptotically flat initial data set}
\begin{Def} \label{intstrasflat2}
A strongly asymptotically flat initial data set is an initial data set 
$(H, \bar{g}, k)$ with: 
\begin{enumerate}
\item $M$ is Euclidean at infinity. \\ 
\item There exists a chart on the neighbourhood of infinity in which the following holds: 
\begin{equation}
\bar{g}_{ij} \ = \ ( 1  \ + \ \frac{2 M}{r} ) \ \delta_{ij} \ + \ o_2 (r^{-1}) \ . 
\end{equation}
\item It is: 
\begin{equation}
k_{ij} \ = \ o_1 (r^{-2}) \ . 
\end{equation}
\end{enumerate}
$M$ denotes the mass. 
\end{Def}
For certain asymptotically flat data sets, the ADM definitions of 
energy $E$, linear momentum $P$ and angular momentum $J$ are well defined and finite. 
Let us write the ADM definitions in the following: 
\begin{Def} (Arnowitt, Deser, Misner (ADM)) \\ \index{ADM}
Let $S_r \ = \ \{ |x| = r  \}$ be the coordinate sphere of radius $r$ and $dS_j$ the 
Euclidean oriented area element of $S_r$. Then we define 
\begin{itemize}
\item Total Energy \index{energy! total energy}
\begin{equation}
E \ = \ \frac{1}{4} \lim_{r \to \infty} \int_{S_r} \sum_{i,j} (\partial_i \bar{g}_{ij} \ - \ 
\partial_j \bar{g}_{ii}) \ dS_j \ , 
\end{equation}
\item Linear Momentum \index{linear momentum!}
\begin{equation}
P^i \ = \ - \frac{1}{2} \lim_{r \to \infty} \int_{S_r}  (k_{ij} \ - \ 
 \bar{g}_{ij} \ tr k) \ dS_j \ , 
\end{equation}
\item Angular Momentum \index{angular momentum!}
\begin{equation}
J^i \ = \ - \frac{1}{2} \lim_{r \to \infty} \int_{S_r} \epsilon_{ijm} x^j (k_{mn} \ - \ 
 \bar{g}_{mn} \ tr k) \ dS_n \ . 
\end{equation}
\end{itemize}
\end{Def}
For strongly asymptotically flat initial data (definition \ref{intstrasflat2}), 
total energy, linear and angular momentum are well defined and conserved. 
Thus, also in the work \cite{sta} of Christodoulou and Klainerman (definition \ref{intSAFCK}), 
all these quantities 
are well defined and conserved. \\ \\ 
In our more general situation (see definition \ref{intAFB}), 
the total energy and the linear momentum are shown to be well defined and conserved. 
We are still within the frame for which R. Bartnik's positive mass theorem applies \cite{bart2}. \\ \\ 
Generally, total energy and linear momentum are well defined and conserved for 
asymptotically flat data sets such that there exists a coordinate system 
in the neighbourhood of infinity in which the following holds 
\bea
\bar{g}_{ij} \ & = & \ \delta_{ij} \ + \ o_2 (r^{- \alpha}) \ , \\ 
k_{ij} \ & = & \ o_1 (r^{-1 - \alpha})  \ \ , \ \ \ \ \alpha \ > \ \frac{1}{2} \ . 
\eea
Let us now discuss the foliation of the spacetime given by $u$. \\ \\ 
The {\itshape optical function} $u$ is a solution of the 
{\itshape Eikonal equation}: 
\be \label{inteik1}
g^{\alpha \beta} \ \frac{\partial u}{\partial x^{\alpha}} \ \frac{\partial u}{\partial x^{\beta}} \ = \ 0 \ . 
\ee
This equation tells us that the level sets $C_u$ of $u$ are null hypersurfaces. \\ \\ 
The {\itshape $(t,u)$ foliations} of the spacetime define a 
codimension $2$ foliation by $2$-surfaces 
\be \label{intSHC1}
S_{t,u} \ = \ H_t \ \cap \ C_u \ , 
\ee
the intersection between $H_t$ (foliation by $t$) and a 
$u$-null-hypersurface $C_u$ (foliation by $u$). 
The 
{\itshape area radius} $r(t,u)$ of $S_{t,u}$ is then defined as: 
\be \label{intdefr12}
r(t,u) \ = \ \sqrt{\frac{\mbox{Area } (S_{t,u})}{4 \pi}} \ \ .
\ee
To construct this optical function $u$, we first choose a $2$-surface $S_{0,0}$, 
diffeomorphic to $S^2$, in $H_0$. We assume the spacetime to have been constructed. Then 
the boundary $\partial J^+ (S_{0,0})$ of the future of $S_{0,0}$ consists of 
an outer and an inner component. They are generated by the congruence of 
outgoing, respectively incoming, null geodesic normals to $S_{0,0}$. 
Now, the zero-level set $C_0$ of $u$ is defined to be this outer component. 
In order to construct all the other level sets $C_u$ for $u \neq 0$, 
we start 
on the last slice $H_{t_*}$ of a spacetime slab. 
We solve on $H_{t_*}$ an equation of motion of surfaces. 
We sketch it later in this paper. 
It forms a crucial part of our work. 
These level sets $C_u$ are also outgoing null hypersurfaces. 
By construction, $u$ is a solution of the Eikonal equation (\ref{inteik1}). \\ \\ \\
Important structures of the spacetime used in the proof 
are coming from a comparison with the Minkowski spacetime. 
Crucial are the 
canonical spacelike foliation, the null structure and the 
conformal group structure. 
As the situation to be studied here is `close' to Minkowksi spacetime, we can use part of its conformal isometry group. 
In \cite{sta}, the authors defined 
the action of the subgroup of the conformal group of Minkowski spacetime 
corresponding to 
the time translations, the scale transformations, the inverted time translations and the 
spatial rotation group $O(3)$. 
For our present proof, we also define the actions for the first three of these, but not for $O(3)$. 
In contrast to \cite{sta}, where the construction of the rotational vectorfields is a major 
part of the proof, we do not work with rotational vectorfields at all. 
Recalling the construction in \cite{sta}, 
once the functions $t$ and $u$ have been fixed, 
the rotation group $O(3)$ takes any given hypersurface $H_t$ onto itself. 
The orbit of $O(3)$ through a given point $p$ is the corresponding 
surface $S_{t,u}$ through $p$. 
The surfaces (\ref{intSHC1}) are the orbits of the rotation group $O(3)$ on $H_t$. 
In our situation, the vectorfields for the time and inverted time translations as well as 
for the scalings supply everything that is needed to obtain the estimates, as we shall see below. 
The group of time translations has already been defined. 
This corresponds to the choice of a canonical time function $t$. 
The integral curves of the generating vectorfield $T$ are the timelike curves orthogonal to the 
hypersurfaces $H_t$, and are parametrized by $t$. For the corresponding group $\{ f_{\tau} \}$ it holds, that 
$f_{\tau}$ is a diffeomorphism of $H_t$ onto $H_{t + \tau}$. 
Further, the vectorfields for the scaling and inverted time translations, that is, $S$ and $K$, respectively, 
are also constructed with the help of the function $u$, as given below. \\ \\ \\
\section{Important Structures and Former Results} 
\label{intsectstructandformerresults*12}
We denote the deformation tensor of $X$ by $\ ^{(X)} \pi$. 
It is given as 
\bea
 \ ^{(X)} \pi_{\alpha \beta} \ & = & \ (\Lie_X g)_{\alpha \beta} \label{intdeformationt*1} \\
-   \ ^{(X)} \pi^{\alpha \beta} \ & = & \ (\Lie_X \ g^{-1})^{\alpha \beta} \ . \label{intdeformationt*2}
\eea
Moreover, a \begin{itshape} Weyl tensor \end{itshape} $W$ is defined to be a 4-tensor that satisfies all the 
symmetry properties of the curvature tensor and in addition is traceless. \\ \\ 
Given a Weyl field $W$ and a vectorfield $X$, the 
\begin{itshape}Lie derivative \end{itshape} 
of $W$ with respect to $X$ is 
not, in general, a Weyl field, for, it has trace. In fact, it is: 
\be
g^{\alpha \gamma}  \ 
(\Lie_X W_{\alpha \beta \gamma \delta}) \ = \  \ ^{(X)} \pi^{\alpha \gamma} \ W_{\alpha \beta \gamma \delta}  \ . 
\ee
In view of this, we define the following 
\begin{itshape}modified Lie derivative: \end{itshape}
\be
\hat{\Lie}_X W \  : =  \ 
\Lie_X W 
\ - \ \frac{1}{2} \ \ ^{(X)} [W] \ + \ \frac{3}{8} \ tr ^{(X)} \pi \ W 
\ee
with 
\be
\ ^{(X)} [W]_{\alpha \beta \gamma \delta} \ : = \  \ ^{(X)} \pi^{\mu}_{\ \alpha} W_{\mu \beta \gamma \delta} \ + \ 
 \ ^{(X)} \pi^{\mu}_{\ \beta} W_{\alpha \mu \gamma \delta} \ + \ 
 \ ^{(X)} \pi^{\mu}_{\ \gamma} W_{\alpha \beta \mu \delta} \ + \ 
 \ ^{(X)} \pi^{\mu}_{\ \delta} W_{\alpha \beta \gamma \mu}  \ . 
\ee
$W$ is said to satisfy the \begin{itshape} Bianchi equation \end{itshape}, if it is: 
\[
D_{[ \alpha} W_{\beta \gamma ] \delta \epsilon} \ = \ 0 . 
\]
To a Weyl field one can associate a tensorial quadratic form, a 4-covariant tensorfield 
which is fully symmetric and trace-free; a generalization of one found previously by 
Bel and Robinson \cite{Bel}. As in \cite{sta} it is called the 
Bel-Robinson tensor: \\ \\ 
\be \label{intQBelRob1}
Q_{\alpha \beta \gamma \delta} \ = \ \frac{1}{2} \ 
(W_{\alpha \rho \gamma \sigma} \ W_{\beta \ \delta}^{\ \rho \ \sigma} \ + \ 
\ ^*W_{\alpha \rho \gamma \sigma } \ \ ^*W_{\beta \ \delta }^{\ \rho \ \sigma } ) \ . 
\ee
It satisfies the following positivity condition:
\be
Q \ (X_1, \ X_2, \ X_3, \ X_4) \ \geq \ 0 
\ee
where $X_1$, $X_2$, $X_3$ and $X_4$ are future-directed timelike vectors. Moreover, 
if $W$ satisfies the Bianchi equations then $Q$ is divergence-free: 
\be \label{divQ=0s}
D^{\alpha} \ Q_{\alpha \beta \gamma \delta} \ = \ 0 \ . 
\ee
Equation (\ref{divQ=0s}) is a property of the Bianchi equations. 
In fact, they are covariant under conformal isometries. 
To be precise, let $\Omega$ be a positive function. Then, if 
$\Phi : M \to M$ is a conformal isometry of the spacetime, i.e., 
\[
\Phi_* g \ = \ \Omega^2 g  \ , 
\]
and if $W$ is a solution, also $\Omega^{-1} \Phi_* W$ is a solution. \\ \\ 
The Bel-Robinson tensor $Q$ is an important tool in our work. 
We shall come back to it. \\ \\ \\
For a long time, the burning question in the Cauchy problem for the EV equations  (\ref{Ricvac0}) 
had been: Does there exist any non-trivial, 
asymptotically flat initial data with complete maximal development? 
In their pioneering work \cite{sta}  ``The global nonlinear stability of the Minkowski space", 
D. Christodoulou and S. Klainerman proved 
global existence and uniqueness of 
such solutions under certain smallness conditions on the initial data. \\ \\ 
But this question has its roots back in the 50s. 
In 1952, Y. Choquet-Bruhat 
focussed the question of local existence and uniqueness of solutions, in GR. 
In \cite{bru} she treated the Cauchy problem for the Einstein equations, 
locally in time, she showed existence and uniqueness of solutions, reducing the Einstein equations to 
wave equations, introducing harmonic coordinates. 
She proved the well-posedeness 
of the local Cauchy problem in these coordinates. 
The local result led to a global theorem proved by Y. Choquet-Bruhat and R. Geroch in \cite{bruger}, 
stating the existence of a unique maximal future development for each given 
initial data set.  \\ \\ 
Regarding the question of completeness or incompleteness of this maximal future development, 
R. Penrose gave an 
answer in his incompleteness theorem 
\cite{icthmrp}, stating that, 
if in the initial data set $(H, \bar{g}, k)$, $H$ is non-compact (but complete), if the positivity condition on the energy holds, 
and $H$ contains 
a closed trapped surface $S$, the boundary of a compact domain in $H$, 
then the corresponding maximal future development is incomplete. 
\begin{Def}
A closed trapped surface $S$ in a non-compact Cauchy hypersurface $H$ is a 
two-dimensional surface in $H$, bounding a compact domain such that 
\[
tr \chi \ < \ 0 \ \ \ \mbox{on } S  \ . 
\]
\end{Def}
The second fundamental form $\chi$ is given below in (\ref{chidef*1}). 
The theorem of Penrose and its extensions by S. Hawking and R. Penrose 
led directly to the question formulated above: 
Is there any non-trivial asymptotically flat initial data whose maximal development is complete? 
The answer was given in the joint work of D. Christodoulou and S. Klainerman \cite{sta}, 
'The global nonlinear stability of the Minkowski space'. 
A rough version of the theorem is stated at the end of subsection \ref{chintsetting***s12}, 
whereas a more precise version is given in theorem \ref{maintheoremCK2}. 
The problem studied by Christodoulou and Klainerman in \cite{sta} was suggested by 
S. T. Yau to Klainerman in 1978. 
D. Christodoulou elaborated the method of \cite{sta} and combined it with new ideas 
in his remarkable work `The Formation of Black Holes in General Relativity' \cite{chrdmay2008}. \\ \\
N. Zipser in \cite{zip} studied the Einstein equations with the energy-momentum tensor being equal to the 
stress-energy tensor of an electro-magnetic field satisfying the Maxwell equations. 
As in \cite{sta}, she considered strongly asymptotically flat initial data 
on a maximal hypersurface. 
She generalized the result of \cite{sta} by proving the global nonlinear stability of the 
trivial solution of the Einstein-Maxwell equations. 
Lately, a proof under stronger conditions 
for the global stability of Minkowski space for the EV equations 
and asymptotically flat Schwarzschild initial data 
was given by H. Lindblad and I. Rodnianski \cite{lindrod1}, \cite{lindrod2}, the latter 
for EV (scalar field) equations. 
They 
worked with a wave coordinate gauge, showing the wave coordinates to be stable globally. 
Concerning the asymptotic behaviour, the results are less precise than the ones 
of Christodoulou and Klainerman in \cite{sta}. Moreover, there are more conditions to be imposed on the 
data than in \cite{sta}. 
There is a variant for the exterior part of the 
proof from \cite{sta} using 
a double-null foliation by S. Klainerman and F. Nicol\`o in \cite{klnico1}. 
Also a semiglobal result was given by H. Friedrich \cite{fried1} 
with initial data on a spacelike hyperboloid.  \\ \\ 
A still open question is: 
What is the sharp critera for 
non-trivial asymptotically flat initial data sets 
to give 
rise to a maximal development that is complete? Or, to what extent can the result of \cite{sta} be generalized? \\ \\ 
The results of \cite{sta}, \cite{zip} and our new result \cite{lydia1}, \cite{lydia2} are much more general 
than the others cited above, as all the other works place stronger conditions on the data. \\ \\ \\ 
\section{Detailed Statement of the Main Results} 
\label{intsectstatementresult*1}
In this section, we provide the most precise version of our main results. 
In order to state our main theorem in full details, we have to introduce 
the corresponding norms. The definitions of these norms can be found in subsection \ref{intnorms*a1}.  \\ \\ 
We recall (\ref{globalsalbQ}) and the global smallness assumption B (\ref{globalsalb1}) 
from the `Introduction'. 
The small positive $\epsilon$ on the right hand side of (\ref{globalsalb1}) has to be chosen 
suitably small later in the proof, depending on other quantities, such that 
the inequalities in the 
main theorem \ref{maintheoremlb*1} hold. \\ 
\begin{The} \label{maintheoremlb*1} {\bf (Main Theorem)} 
Any asymptotically flat, maximal initial data set (AFB)  
of the form given in definition \ref{intAFB} 
in the `Introduction' 
satisfying the global smallness assumption B stated in the `Introduction', 
inequality (\ref{globalsalb1}), 
leads to a unique, globally hyperbolic, smooth and geodesically complete solution of the EV equations, 
foliated by the level sets of a maximal time function $t$, defined for all $t \geq -1$. 
Moreover, there exists a global, smooth optical function $u$, that is a solution of the 
Eikonal equation defined everywhere in the exterior region 
$r \geq \frac{r_0}{2}$, with $r_0(t)$ denoting the radius 
of the $2$-surface $S_{t, 0}$ of intersection between the hypersurfaces $H_t$ and 
a fixed null cone $C_0$ with vertex at a point on $H_{-1}$. 
With respect to this foliation the following holds: 
\bea
\ ^ e \mathcal{R}_{[1]} , \  
\ ^ e \mathcal{K}_{[2]} , \  
\ ^ e \mathcal{O}_{[2]} , \ 
\ ^ e \mathcal{L}_{[2]} 
\ \ \leq \ \ \epsilon_0   \label{maintheoremres*1} \\ 
\ ^ e \mathcal{K}_{0}^{\infty} , \  
\ ^ e \mathcal{O}_{0}^{\infty} , \  
\ ^ e \mathcal{L}_{0}^{\infty}  
\ \ \leq \ \ \epsilon_0   \label{maintheoremres*2} \ \ . 
\eea
Moreover, in the complement of the exterior region, the following holds: 
\bea
\ ^i \mathcal{R}_{[1]} , \  
\ ^ i \mathcal{K}_{[2]} , \  
\ ^ i \mathcal{L}_{[2]} 
\ \ \leq \ \ \epsilon_0   \label{maintheoremres*3} \\ 
\ ^ i \mathcal{K}_{0}^{\infty} , \  
\ ^ i \mathcal{L}_{0}^{\infty}  
\ \ \leq \ \ \epsilon_0   \label{maintheoremres*4} \ \ . 
\eea
The strict inequalities hold for $t=0$ with $\epsilon$ on the right hand sides. \\ \\ 
The norms are given in subsection \ref{intnorms*a1}. 
\[\]
\[\]
\end{The}
In the next section, we are going to state the main steps of the proof and to 
explain the bootstrap argument in details. \\ \\ 
We recall that in the global smallness assumption B (\ref{globalsalb1})  
the initial data has to be smaller than a sufficiently small positive $\epsilon$. 
Later in the proof, this $\epsilon$, has to be taken suitably small, depending 
on other quantities. 
In the bootstrap assumptions BA0-BA2 ((\ref{BA0})-(\ref{BA2})) and in (\ref{addass**la1}) 
the considered quantities have to be smaller than 
a small positive $\epsilon_0$. 
We estimate the main quantities at times $t$ by their values at $t=0$, which 
are controlled by inequalities with $\epsilon$ on their right hand sides. 
Then, choosing $\epsilon$ sufficiently small, the right hand sides of these inequalities can be 
made strictly smaller than $\epsilon_0$ from the bootstrap assumptions. 
The bootstrap argument is explained in details in the next section. \\ \\ 
Our main theorem \ref{maintheoremlb*1} 
provides existence and uniqueness of solutions under the relaxed assumptions 
(AFB, see (\ref{globalsalb1})) as well as it describes the asymptotic behaviour as 
precisely as it is possible under these relaxed assumptions. 
Compared to the result \cite{sta}, main theorem 10.2.1, p. 298,  
by Christodoulou and Klainerman, 
(we give one version of their main theorem in theorem \ref{maintheoremCK2}), 
we impose less on our initial data. That is, we assume one less power of $r$ decay of the data at infinity and 
one less derivative to be controlled. \\ \\ 
%
%
%
%
%
%
%
%
\section{Crucial Steps of the Proof of the Main Theorem - Bootstrap Argument}
\label{subchcrucialstepsproofBA***12}
Our proof consists of one large 
{\itshape bootstrap argument}, 
containing other arguments of the same type but at different levels. \\ \\ 
First, we give the {\bf main steps of the proof of our main result.} They can be summarized as follows: 
\begin{itemize}
\item[1.] {\itshape Energy}. Estimate a {\itshape quantity} 
$\mathcal{Q}_1(W)$ , which is an 
integral over $H_t$ involving the Bel-Robinson tensor $Q$ of the spacetime curvature $W$ 
and of the Lie derivatives of $W$ as below. 
At time $t$, this quantity $\mathcal{Q}_1(W)$  
can be calculated by its value at $t=0$ and an integral from $0$ to $t$, which 
both are controlled. 
We use the vectorfields $T$ time translations, $S$ scaling, $K$ inverted time translations and 
$\bar{K} = K + T$. (See below). 
$\mathcal{Q}_1(W)$ is given by: 
\bea
\mathcal{Q}_1(W) \ & = & \ 
Q_0  \ + \ Q_1   \label{intQ1W12}
\eea
with $Q_0$ and $Q_1$ being the following integrals, 
\bea
Q_0 (t) \ & = & \ 
\int_{H_t} \ Q \  (W) \ (\bar{K}, T , T, T)  \label{intQ0W12} \\ 
Q_1 (t) \ & = & \ 
\int_{H_t} \ Q \ (\hat{\Lie}_S W) \ (\bar{K}, T , T, T)  \nonumber \\ 
& & \ + \ 
\int_{H_t} \ Q \ (\hat{\Lie}_T W) \ (\bar{K}, \bar{K} , T, T) \ .  \label{intQ1W123}
\eea
\item[2.] The components of the \begin{itshape}Weyl tensor \end{itshape}$W$ 
are estimated by a 
{\itshape comparison argument} with $\mathcal{Q}_1(W)$. 
The estimates for the Weyl tensor $W$ rely heavily on the fact that $W$ satisfies the Bianchi equations. 
\item[3.] The \begin{itshape}geometric \end{itshape}quantities are estimated from 
\begin{itshape}curvature assumptions \end{itshape}using the optical structure equations, 
elliptic estimates, 
evolution equations, and tools like Sobolev inequalities. 
\end{itemize}
The bootstrapping allows us to go from local to global. The whole procedure can mainly be 
split into three parts. 
\begin{itemize}
\item \begin{itshape}Bootstrap assumptions: \end{itshape}Initial assumptions on the 
main geometric quantities of the two foliations, 
i.e. $\{ H_t \}$ and $\{ C_u \}$. 
\item \begin{itshape}Local existence theorem: \end{itshape} 
It guarantees the local existence of a unique solution and the preservation of the 
asymptotic behaviour in space 
of the metric, the second fundamental form and the curvature. 
\item \begin{itshape}Bootstrap argument: \end{itshape} 
Together with the evolution equations it yields the 
global existence of a unique solution as well as it shows the asymptotic behaviour as above to be preserved. 
\end{itemize}
Now, we are going to state and to explain the crucial steps of the proof of the main theorem, that is, 
of the 
{\itshape bootstrap argument}.   
\index{bootstrap argument!} \\ \\ 
Before that, let us say a few words about the geometric quantities in point three. 
In order to estimate the geometric quantities of point three above in the framework of a 
bootstrap argument, one assumes that the curvature components satisfy suitable bounds. 
To see how this is done, let us consider the affine foliation $\{ S_s \}$ of $C$. 
Here, the generating vectorfield $L$ of $C$ is geodesic, and $s$ is the corresponding affine parameter function. 
We shall now give the idea for the 
case of the second fundamental form $\chi$ of $S_s$ relative to $C$ 
\be \label{chidef*1}
\chi (X, Y) \ = \ g (D_X L, Y) 
\ee
for any pair of vectors $X,Y \in T_pS_s$. 
To estimate $\chi$, we first split it into its trace $tr \chi$ and traceless part $\hat{\chi}$. Then, 
one estimates $\chi$ from the propagation equation 
\be \label{dtrchi}
\frac{\partial tr \chi }{\partial s} \ + \ \frac{1}{2} \ (tr \chi)^2 \ + \ \mid \hat{\chi} \mid^2  
\ = \ 0   
\ee
and the elliptic system on each section $S_s$ of $C$ (the Codazzi equations) 
\be \label{chihatfa}
\dlap \ \hat{\chi}_a \ = \ \frac{1}{2} \ d_a tr \chi  \ + \ f_a 
\ee
where 
\[
f_a \ \ \mbox{involves curvature} . 
\]
Assuming estimates for the spacetime curvature
on the right hand side of (\ref{chihatfa}) 
yields estimates for the 
quantities controlling the geometry of $C$ as described by its foliation $\{ S_s \}$. 
This is discussed in more details below. \\ \\ 
\subsection{Local Existence Theorem}
Relying on the 
{\itshape local existence theorem}, \index{local existence theorem!} 
\cite{sta}, statement theorem 10.2.2: p. 299/300, 
proof 10.2.2: p. 304 - 310, 
we show global existence of 
a unique, globally hyperbolic, smooth and geodesically complete solution of the Einstein-vacuum equations, coming 
from initial data stated in definition \ref{intAFB} (AFB) and inequality \ref{globalsalb1} 
(global smallness assumption B). 
The local existence theorem, is stated and proven in \cite{sta} for their problem, 
and it also holds in our case. 
It requires the second fundamental form $k$ to be in $L^{\infty}$, which is satisfied in our situation. 
The proof of the local existence theorem in \cite{sta} mainly uses the ideas developed in the proof 
of the well-known existence result of Choquet-Bruhat \cite{bru} and modifies them simply. 
In \cite{sta}, the authors formulate the conditions in the local existence theorem, 
according to their situation imposing 
$Ric(\bar{g}_0) \in H_{2,1}(H, \bar{g}_0)$ and $k \in H_{3,1}(H, \bar{g}_0)$, 
while we impose 
$Ric(\bar{g}_0) \in H_{1,1}(H, \bar{g}_0)$ and $k \in H_{2,0}(H, \bar{g}_0)$. 
The proof still holds in the same way, as it only requires $k$ in $L^{\infty}$, which in our situation 
is true. \\ \\ 
\subsection{Norms} 
\label{intnorms*a1} 
We define the norms as they appear in the main theorem \ref{maintheoremlb*1} and 
as we use them in subsection \ref{norms*BA**subsection*1}. \\ \\ 
We consider a spacetime slab 
$\bigcup_{t \in [0, t_*]} H_t$. In this section, we assume that this slab is foliated by a  
maximal time function $t$ and by an optical function $u$. 
In what follows, we shall introduce the basic norms for the curvature $R$, the second fundamental form 
$k$, the lapse function $\phi$ and the components $\chi, \zeta, \omega$ of the Hessian of $u$. 
In most of the definitions below, we follow the notation of \cite{sta}. \\ \\ 
Let $V$ be a vectorfield tangent to $S$. Then we define the norms on $S$: 
\bea
\parallel V \parallel_{p, S} (t,u) \ & = & \ 
\Big(
\int_{S_{t,u}} \mid V \mid^p \ d \mu_{\gamma} \Big)^{\frac{1}{p}}  \ \ , \ \ \mbox{ for } 1 \ \leq \ p \ < \infty 
\label{mainthmnormsS*1} \\ 
 \ & = & \ 
\sup_{S_{t,u}} \mid V \mid \ \  , \ \ \mbox{ for } p \ = \ \infty \ \ . 
\label{mainthmnormsS*2}  
\eea
Sometimes, we also denote these norms by $\mid V \mid_{p, S} (t,u)$. \\ \\ 
The following norms are stated for the interior and the exterior regions of each hypersurface $H_t$. \\ \\ 
The 
\begin{itshape}interior region $I$\end{itshape}, also denoted by $H_t^i$,  
are all the points in $H_t$ for which 
\[
r \ \leq \ \frac{r_0(t)}{2} \ . 
\]
The 
\begin{itshape}exterior region $U$\end{itshape}, also denoted by $H_t^e$,  
are  all the points in $H_t$ for which 
\[
r \ \geq \ \frac{r_0(t)}{2} \ . 
\]
Here, $r_0(t)$ is the value of $r$ corresponding to the area of $S_{t,0}$, the surface of intersection between 
$C_0$ and $H_t$. And $u_1(t)$ is the value of $u$ corresponding to $r_0(t)/2$. \\ \\  
Now, we introduce 
\bea
\parallel V  \parallel_{p, i}  \ & = & \ 
\Big( \int_{H_t^i} \mid V \mid^p \Big)^{\frac{1}{p}}  \ \ , \ \ \mbox{ for } 1 \ \leq \ p \ < \infty 
\label{mainthmnormsI*1}  \\ 
\parallel V  \parallel_{\infty, i}  \ & = & \ 
\sup_{H_t^i} \mid V \mid  \label{mainthmnormsI*2}  \\ 
\nonumber \\ 
\parallel V \parallel_{p, e} (t) \ & = & \ 
\Big( \int_{H_t^e} \mid V \mid^p \Big)^{\frac{1}{p}}  \ \ , \ \ \mbox{ for } 1 \ \leq \ p \ < \infty 
\label{mainthmnormsU*1}  \\ 
\parallel V  \parallel_{\infty, e} (t)  \ & = & \ 
\sup_{H_t^e} \mid V \mid  \label{mainthmnormsU*2}  \ \ . 
\eea
Sometimes, we denote the norms ((\ref{mainthmnormsI*1})-(\ref{mainthmnormsU*2})) also 
by $\mid V \mid_{p, I}$, $\mid V \mid_{p, U} (t)$ or 
$\parallel V \parallel_{p, I}$, $\parallel V \parallel_{p, U} (t)$ or 
$\mid V \mid_{p, i}$, $\mid V \mid_{p, e} (t)$, respectively.  \\ \\ 
\subsubsection{Norms for the Curvature Tensor $R$}
\label{norms*curvature**subsection*1}
We are now going to define the norms 
$\overline{\mathcal{R}_0} \ (W)$ and 
$\overline{\mathcal{R}_{1}} \ (W)$ as well as 
$\mathcal{R}_0(W)$ and $\mathcal{R}_1(W)$ 
for the null curvature components. \\ \\ 
The norms are stated for the interior and exterior regions of the hypersurface $H_t$. \\ \\ 
{\bf Norms}
$\mathbf{\overline{\mathcal{R}_{0}} \ (W)}$ {\bf and} 
$\mathbf{\overline{\mathcal{R}_{1}} \ (W)}$: \\ \\ 
On each slice $H_t$, we denote by 
$\overline{\mathcal{R}_{q}} \ (W)$ the following maximum: 
\be \label{XcompnormsR*1}
\overline{\mathcal{R}_{q}} \ (W) \ = \ \max 
\big( \overline{\ ^e \mathcal{R}_q}(W) , \  \overline{\ ^i \mathcal{R}_q} (W)  \big) \ \ \ \quad ,  
\ \ q \ = \ 0 , \ 1 \ . 
\ee
By $\overline{\ ^e \mathcal{R}_q}(W)$ we denote the exterior and by 
$\overline{\ ^i \mathcal{R}_q} (W)$ the interior $L^2$-norms of the curvature given 
as follows: In the interior region $I$, we define for $q = 0, 1$, 
\be \label{XcompnormsR*2}
\overline{\ ^i \mathcal{R}_q} \ = \ r_0^{1 + q} \ \parallel D^q \ W \parallel_{2,I} \ . 
\ee
Then, one sets 
\bea
\overline{\ ^i \mathcal{R}_{[0]}} \ & = & \ \overline{\ ^i \mathcal{R}_0} \label{XcompnormsolR**2*1}  \\ 
\overline{\ ^i \mathcal{R}_{[1]}} \ & = & \ \overline{\ ^i \mathcal{R}_{[0]}} \ + \ 
\overline{\ ^i \mathcal{R}_1}  \label{XcompnormsolR**2*2}  \ \ . 
\eea
We define the exterior norms $\overline{\ ^e \mathcal{R}_0}(W)$ and $\overline{\ ^e \mathcal{R}_1}(W)$ to be  
\bea
\overline{\ ^e\mathcal{R}_0} \ (W)^2 \ & = & \ 
\int_{U} \tau_-^2  \mid \underline{\alpha} \mid^2 \ + \ 
\int_{U}  r^2 \mid \underline{\beta} \mid^2 \ + \ 
\int_{U} r^2 \mid \rho \mid^2 \ + \ 
\int_{U} r^2 \mid \sigma \mid^2  \nonumber \\ 
& & 
\ + \ 
\int_{U} r^2 \mid \beta \mid^2 \ + \ 
\int_{U} r^2 \mid \alpha \mid^2 
\eea
and 
\bea
\overline{\ ^e\mathcal{R}_{1}} \ (W)^2 \ & = & \ 
\int_{U} \tau_-^2 r^2 \mid \nlap \underline{\alpha} \mid^2 \ + \ 
\int_{U}  r^4 \mid \nlap \underline{\beta} \mid^2 \ + \ 
\int_{U} r^4 \ \mid \nlap \rho \mid^2 \ + \ 
\int_{U} r^4 \ \mid \nlap \sigma \mid^2  \nonumber \\
& & 
\ + \ 
\int_{U} r^4 \mid \nlap \beta \mid^2 \ + \ 
\int_{U} r^4 \mid \nlap \alpha \mid^2   \nonumber \\ 
& & 
\ + \ 
\int_{U} \tau_-^4 \mid \nlap_{N} \underline{\alpha} \mid^2 \ + \ 
\int_{U} \tau_-^2 r^2 \mid \nlap_{N}  \underline{\beta} \mid^2 \ + \ 
\int_{U} r^4 \mid \nlap_{N} \rho \mid^2 \ + \ 
\int_{U} r^4 \mid \nlap_{N} \sigma \mid^2  \nonumber \\
& & 
\ + \ 
\int_{U} r^4 \mid \nlap_{N} \beta \mid^2 \ + \ 
\int_{U} r^4 \mid \nlap_{N} \alpha \mid^2 \ . \\ 
& & \nonumber
\eea 
We refer to the norms of the components of $R$ by the formulas: for $q = 0, 1$: 
\beas
 \overline{\ ^e\mathcal{R}_q} (\underline{\alpha}) \ & = & \ 
\parallel \tau_- r^q \nlap^q  \underline{\alpha} \parallel_{2, e} \\ 
 \overline{\ ^e\mathcal{R}_q} (\alpha)  \ & = & \ 
\parallel r^{q+1} \nlap^q \alpha \parallel_{2, e}  \\ 
\cdots 
\eeas
and correspondingly for the remaining components. \\ \\ 
Denote 
$\underline{\alpha}_N = \nlap_N \underline{\alpha}$, 
$\alpha_N = \nlap_N \alpha$ and correspondingly for the other curvature components. 
Then, we set 
\bea
 \overline{\ ^e\mathcal{R}_0} [\underline{\alpha}] \ & = & \ 
 \overline{\ ^e\mathcal{R}_0} (\underline{\alpha})  \nonumber \\  
 \overline{\ ^e\mathcal{R}_1} [\underline{\alpha}] \ & = & \ 
\Big( \ 
 \overline{\ ^e\mathcal{R}_1} (\underline{\alpha})^2 \ + \ 
 \overline{\ ^e\mathcal{R}_0} (\underline{\alpha}_N)^2 
\ \Big)^{\frac{1}{2}}  \label{XcompnormsolR**squares*1}
\eea
Similarly as in (\ref{XcompnormsolR**squares*1}), we proceed with all the other null components of the curvature. 
This allows us now to define for $q = 0, 1$ the following: 
\be \label{XcompnormsolR**squares*2}
 \overline{\ ^e\mathcal{R}_q} \ = \ 
\Big( \ 
 \overline{\ ^e\mathcal{R}_q} [\underline{\alpha}]^2 \ + \ 
  \overline{\ ^e\mathcal{R}_q} [\underline{\beta}]^2 \ + \ \cdots \ + \ 
 \overline{\ ^e\mathcal{R}_q} [\alpha]^2  
 \ \Big)^{\frac{1}{2}}   \ \ . 
\ee 
Then, one sets 
\bea 
\overline{\ ^e\mathcal{R}_{[0]}} \ & = & \ \overline{\ ^e\mathcal{R}_0}  \label{XcompnormsolR**2*3}  \\  
\overline{\ ^e\mathcal{R}_{[1]}} \ & = & \ \overline{\ ^e\mathcal{R}_{[0]}}  \ + \ 
\overline{\ ^e\mathcal{R}_1}  \ \ .  \label{XcompnormsolR**2*3}  \\  
\nonumber 
\eea 
\[\]
\[\]
{\bf Norms}
$\mathbf{\mathcal{R}_0(W)}$ {\bf and} $\mathbf{\mathcal{R}_1(W)}$: \\ \\ 
Similarly as above, 
we define on each slice $H_t$ the quantity 
$\mathcal{R}_{q} \ (W)$ as the maximum: 
\be \label{XcompnormsR**1}
\mathcal{R}_{q} \ (W) \ = \ \max 
\big( \ ^e \mathcal{R}_q (W) , \  \ ^i \mathcal{R}_q (W)  \big) \ \ \ \quad ,  
\ \ q \ = \ 0 , \ 1 \ . 
\ee
By $\ ^e \mathcal{R}_q(W)$ we denote the exterior and by 
$\ ^i \mathcal{R}_q (W)$ the interior $L^2$-norms of the curvature given 
as follows: In the interior region $I$, we define for $q = 0, 1$, 
\be \label{XcompnormsR**2}
\ ^i \mathcal{R}_q \ = \ r_0^{1 + q} \ \parallel D^q \ W \parallel_{2,I} \ . 
\ee
Then, one sets 
\bea
\ ^i \mathcal{R}_{[0]} \ & = & \ \ ^i \mathcal{R}_0  \label{XcompnormsR**2*1}  \\ 
\ ^i \mathcal{R}_{[1]} \ & = & \ \ ^i \mathcal{R}_{[0]} \ + \   \ ^i \mathcal{R}_1  \label{XcompnormsR**2*2}  \ \ . 
\eea
The exterior norms $\ ^e \mathcal{R}_0$ and $\ ^e \mathcal{R}_1$ we define as follows: 
\bea
\ ^e\mathcal{R}_0 \ (W)^2 \ & = & \ 
\int_{U} \tau_-^2  \mid \underline{\alpha} \mid^2 \ + \ 
\int_{U}  r^2 \mid \underline{\beta} \mid^2 \ + \ 
\int_{U} r^2 \mid \rho \mid^2 \ + \ 
\int_{U} r^2 \mid \sigma \mid^2  \nonumber \\ 
& & 
\ + \ 
\int_{U} r^2 \mid \beta \mid^2 \ + \ 
\int_{U} r^2 \mid \alpha \mid^2  \label{XnormR0out}
\eea
and 
\bea
\ ^e\mathcal{R}_{1} \ (W)^2 \ & = & \ 
\int_{U} \tau_-^2 r^2 \mid \nlap \underline{\alpha} \mid^2 \ + \ 
\int_{U}  r^4 \mid \nlap \underline{\beta} \mid^2 \ + \ 
\int_{U} r^4 \ \mid \nlap \rho \mid^2 \ + \ 
\int_{U} r^4 \ \mid \nlap \sigma \mid^2  \nonumber \\
& & 
\ + \ 
\int_{U} r^4 \mid \nlap \beta \mid^2 \ + \ 
\int_{U} r^4 \mid \nlap \alpha \mid^2   \nonumber \\ 
& & 
\ + \ 
\int_{U} \tau_-^4 \mid  \underline{\alpha}_3 \mid^2 \ + \ 
\int_{U} \tau_-^4 \mid  \underline{\alpha}_4 \mid^2 \ + \ 
\int_{U} \tau_-^2 r^2 \mid   \underline{\beta}_3 \mid^2 \ + \ 
\int_{U} r^4 \mid   \underline{\beta}_4 \mid^2  \nonumber \\ 
& & 
\ + \ 
\int_{U} r^4 \mid  \rho_3 \mid^2 \ + \ 
\int_{U} r^4 \mid  \rho_4 \mid^2 \ + \ 
\int_{U} r^4 \mid  \sigma_3 \mid^2  \ + \ 
\int_{U} r^4 \mid  \sigma_4 \mid^2 \nonumber \\
& & 
\ + \ 
\int_{U} r^4 \mid  \beta_3 \mid^2 \ + \ 
\int_{U} r^4 \mid  \beta_4 \mid^2 \ + \ 
\int_{U} r^4 \mid  \alpha_3 \mid^2 \ + \ 
\int_{U} r^4 \mid  \alpha_4 \mid^2 \ .  \label{XnormR1out}
\eea
We refer to the norms of the components of $R$ by the formulas: for $q = 0, 1$: 
\beas
 \ ^e \mathcal{R}_q  (\underline{\alpha}) \ & = & \ 
\parallel \tau_- r^q \nlap^q  \underline{\alpha} \parallel_{2, e} \\ 
 \ ^e \mathcal{R}_q 
(\alpha)  \ & = & \ 
\parallel r^{q+1} \nlap^q \alpha \parallel_{2, e}  \\ 
\cdots 
\eeas
and correspondingly for the remaining components. \\ \\ 
Then, we set 
\bea
\ ^e \mathcal{R}_{0} [\underline{\alpha}] \ & = & \ 
\ ^e \mathcal{R}_{0} (\underline{\alpha})  \nonumber \\ 
\ ^e \mathcal{R}_{1} [\underline{\alpha}] \ & = & \ 
\Big( \ 
\ ^e \mathcal{R}_{1} (\underline{\alpha})^2 \ + \ 
\ ^e \mathcal{R}_{0} (\underline{\alpha}_3)^2 \ + \ 
\ ^e \mathcal{R}_{0} (\underline{\alpha}_4)^2 
\ \Big)^{\frac{1}{2}}  \label{XompnormsR**squares*1}  
\eea
Similarly, this is done for all the other null components of the curvature. 
Thus, we define the following for $q = 0, 1$: 
\be  \label{XcompnormsR**squares*2}  
\ ^e \mathcal{R}_{q} \ = \ 
\Big( \ 
\ ^e \mathcal{R}_{q} [\underline{\alpha}]^2 \ + \ 
\ ^e \mathcal{R}_{q} [\underline{\beta}]^2 \ + \ \cdots \ + \ 
\ ^e \mathcal{R}_{q} [\alpha]^2 
\ \Big)^{\frac{1}{2}}  \ \ . 
\ee
Then, one sets 
\bea
\ ^e \mathcal{R}_{[0]} \ & = & \ \ ^e \mathcal{R}_0 \label{XcompnormsR**2*3}  \\  
\ ^e \mathcal{R}_{[1]} \ & = & \ \ ^e \mathcal{R}_{[0]} \ + \  \ ^e \mathcal{R}_1  \label{XcompnormsR**2*4}  \ \ . 
\eea
\[\]
\subsubsection{Norms for the Second Fundamental Form $k$ of the $t$-Foliation} 
\label{norms*secondfundffk**subsection*1}
First, let us define 
$ \ ^i \mathcal{K}_q^p$ and 
$ \ ^e \mathcal{K}_q^p$ to be the interior and exterior weighted $L^p$-norms of the $q$-covariant 
derivatives of the components of the second fundamental form $k$. The quantity 
$\mathcal{K}_q^p$ we then define as follows: 
\be \label{compnormsK**1}
\mathcal{K}_{q}^p \ = \ \max 
\big( \ ^ i \mathcal{K}_q^p  , \  \ ^e \mathcal{K}_q^p   \big) \ \ \ \quad ,  \ \ \ q \ = \ 0 , \ 1 , \ 2 \  \  
\mbox{ and } 
\ 1 \ \leq \ p \ < \  \infty \ \ . 
\ee
Correspondingly, we have 
\be \label{compnormsK**2}
\mathcal{K}_{0}^{\infty} \ = \ \max 
\big( \ ^ i \mathcal{K}_0^{\infty}  , \  \ ^e \mathcal{K}_0^{\infty}   \big) 
\ee
The interior norms $\ ^ i \mathcal{K}_q^p$ in (\ref{compnormsK**1}) are given by: 
\be \label{compnormsK**3}
 \ ^ i \mathcal{K}_q^p  \ = \ 
r_0^{1 + q - \frac{2}{p}} \parallel D^q k \parallel_{p, i} \ \ . 
\ee
In view of the exterior norms $ \ ^e \mathcal{K}_q^p$, let us remind ourselves that 
the second fundamental form $k$ relative to the radial foliation of $u$ on $H_t$ decomposes into 
\bea
k_{NN} \ & = & \ \delta  \nonumber \\
k_{AN}  \ & = & \ \epsilon_A  \nonumber \\ 
k_{AB} \ & = & \ \eta_{AB}  \ \ .  \label{mainthmnormskcops*1}  
\eea
In addition, $\eta$ decomposes into its trace $tr \eta = - \delta$ and its traceless part $\hat{\eta}$. 
Let us also introduce the following notation: 
\bea
\delta_4 \ & = & \ 
\cDlap_4 \ \delta  \nonumber \\ 
\epsilon_4 \ & = & \ 
\cDlap_4 \ \epsilon  \nonumber \\ 
\hat{\eta}_4 \ & = & \ 
\cDlap_4 \ \hat{\eta} \ + \ \frac{1}{2} \ tr \chi \ \hat{\eta} \nonumber \\ 
\delta_3 \ & = & \ 
\cDlap_3 \ \delta  \nonumber \\ 
\epsilon_3 \ & = & \ 
\cDlap_3 \ \epsilon \nonumber \\ 
\hat{\eta}_3 \ & = & \ 
\cDlap_3 \ \hat{\eta}  \ + \ \frac{1}{2} \ tr \underline{\chi} \ \hat{\eta}   \label{mainthmnormskcops*2} 
\eea
Then, we set 
\bea
\ ^e \mathcal{K}_q^p (\delta) \ & = & \ 
\parallel r^{(\frac{3}{2} - \frac{3}{p} + q )} \ \nlap^q \delta \parallel_{p, e}  \nonumber \\ 
 \ ^e \mathcal{K}_q^p (\epsilon) \ & = & \ 
\parallel r^{(\frac{3}{2} - \frac{3}{p} + q )} \ \nlap^q \epsilon \parallel_{p, e}  \nonumber \\ 
\ ^e \mathcal{K}_q^p (\hat{\eta}) \ & = & \ 
\parallel r^{(\frac{3}{2} - \frac{3}{p} + q )} \ 
\ \nlap^q \hat{\eta} \parallel_{p, e}  
\label{mainthmnormskcops*3} \\ 
\nonumber \\ 
\ ^e \mathcal{K}_{q+1}^p (\delta_4) \ & = & \ 
\parallel r^{(\frac{5}{2} - \frac{3}{p} + q )} \  
\nlap^q  \delta_4 
\parallel_{p, e}    \nonumber \\ 
\ ^e \mathcal{K}_{q+1}^p (\delta_3) \ & = & \ 
\parallel r^{(\frac{3}{2} - \frac{3}{p} + q )} \tau_-^{(\frac{3}{2} - \frac{1}{p})} \  
\nlap^q  \delta_3 
\parallel_{p, e}    \nonumber \\ 
\ ^e \mathcal{K}_{q+1}^p (\epsilon_4) \ & = & \ 
\parallel r^{(\frac{5}{2} - \frac{3}{p} + q )} \  
\nlap^q  \epsilon_4 
\parallel_{p, e}    \nonumber \\ 
\ ^e \mathcal{K}_{q+1}^p (\epsilon_3) \ & = & \ 
\parallel r^{(\frac{3}{2} - \frac{3}{p} + q )} \tau_-^{(\frac{3}{2} - \frac{1}{p})} \  
\nlap^q  \epsilon_3 
\parallel_{p, e}    \nonumber \\ 
\ ^e \mathcal{K}_{q+1}^p (\hat{\eta}_4) \ & = & \ 
\parallel r^{(\frac{5}{2} - \frac{3}{p} + q )} \  
\nlap^q  \hat{\eta}_4 
\parallel_{p, e}    \nonumber \\ 
\ ^e \mathcal{K}_{q+1}^p (\hat{\eta}_3) \ & = & \ 
\parallel r^{(1 - \frac{2}{p} + q )}  \tau_-^{(\frac{3}{2} - \frac{1}{p})}   
\nlap^q  \hat{\eta}_3 
\parallel_{p, e}  
\label{mainthmnormskcops*4}  
\eea
Next, one sets 
\bea
\ ^e \mathcal{K}_0^p [\delta] \ & = & \ 
\ ^e \mathcal{K}_0^p (\delta)  \nonumber \\ 
 \ ^e \mathcal{K}_1^p [\delta] \ & = & \ 
\ ^e \mathcal{K}_1^p (\delta) \ + \ \ ^e \mathcal{K}_1^p (\delta_3) \ + \  \ ^e \mathcal{K}_1^p (\delta_4)  
\nonumber \\ 
 \ ^e \mathcal{K}_2^p [\delta] \ & = & \ 
\ ^e \mathcal{K}_2^p (\delta) \ + \ \ ^e \mathcal{K}_2^p (\delta_3) \ + \  \ ^e \mathcal{K}_2^p (\delta_4)  \ \ , 
\label{mainthmnormskcops*5}   
\eea
and correspondingly we do this for $\epsilon$ and $\hat{\eta}$. Then it 
is for all $q = 0, 1, 2$: 
\be
\ ^e \mathcal{K}_q^p \ = \ 
\ ^e \mathcal{K}_q^p [\delta] \ + \ 
\ ^e \mathcal{K}_q^p [\epsilon] \ + \ 
\ ^e \mathcal{K}_q^p [\hat{\eta}] \ . 
\ee
For the case $p=2$, that will be used later on, 
we write simply 
\[
\mathcal{K}_q \ = \ \mathcal{K}_q^2 \ . 
\]
Thus, we define the basic spacetime norms for $k$ as follows: 
\bea
\ ^i \mathcal{K}_{[0]} \ & = & \ 
\ ^i \mathcal{K}_0  \label{mainthmnormskcops*6} \\ 
\ ^i \mathcal{K}_{[1]} \ & = & \ 
\ ^i \mathcal{K}_{[0]} \ + \ 
\ ^i \mathcal{K}_1  \label{mainthmnormskcops*7} \\ 
\ ^i \mathcal{K}_{[2]} \ & = & \ 
\ ^i \mathcal{K}_{[1]} \ + \ 
\ ^i \mathcal{K}_2  \label{mainthmnormskcops*8} \\ 
\ ^e \mathcal{K}_{[0]} \ & = & \ 
\ ^e \mathcal{K}_0  \label{mainthmnormskcops*9} \\ 
\ ^e \mathcal{K}_{[1]} \ & = & \ 
\ ^e \mathcal{K}_{[0]} \ + \ 
\ ^e \mathcal{K}_1  \label{mainthmnormskcops*10} \\ 
\ ^e \mathcal{K}_{[2]} \ & = & \ 
\ ^e \mathcal{K}_{[1]} \ + \ 
\ ^e \mathcal{K}_2  \label{mainthmnormskcops*11} \\ 
\mathcal{K}_{[q]} \ & = & \ \ ^i \mathcal{K}_{[q]} \ + \ \ ^e \mathcal{K}_{[q]} \label{mainthmnormskcops*12} 
\\ \nonumber \\ \nonumber 
\eea
Correspondingly, we define 
$\overline{ \ ^i \mathcal{K}_q^p}$ and 
$\overline{ \ ^e \mathcal{K}_q^p}$ to be the interior and exterior weighted $L^p$-norms as given below. 
Then 
$\overline{\mathcal{K}_q^p}$ we define as follows: 
\be \label{compnormsKol**1}
\overline{\mathcal{K}_{q}^p} \ = \ \max 
\big( \overline{\ ^ i \mathcal{K}_q^p}  , \ \overline{ \ ^e \mathcal{K}_q^p} \big) \ \ \ \quad ,  
\ \ \ q \ = \ 0 , \ 1 , \ 2 \  \  
\mbox{ and } 
\ 1 \ \leq \ p \ < \  \infty \ \ . 
\ee
Correspondingly, we have 
\be \label{compnormsKol**2}
\overline{\mathcal{K}_{0}^{\infty}} \ = \ \max 
\big( \overline{\ ^ i \mathcal{K}_0^{\infty}}  , \  \overline{\ ^e \mathcal{K}_0^{\infty} }  \big) 
\ee
Now, in the interior region $I$ 
we define the following norms for $k$: 
\be \label{compnormsKol**3}
\overline{\ ^i \mathcal{K}_q^p} \ \ = \ \
r_0^{1 + q - \frac{2}{p}} \parallel D^q k \parallel_{p, i} \ \ . 
\ee
On the other hand, in the exterior region $U$, 
we define the following norms: 
\bea
\overline{\ ^e \mathcal{K}_q^p} (\delta) \ & = & \ 
\Big{\{}
\parallel 
r^{(\frac{3}{2} - \frac{3}{p} + q )} \ 
\nlap^q \delta \parallel^p_{p, e} 
\ + \ 
\sum_{i+j = q, \ j \geq 1}  \parallel r^{ (2 - \frac{2}{p} + i ) } \ (\tau_-)^{1 - \frac{2}{p} + j - 1}  \ 
\nlap^i \nabla^j_N \delta 
\parallel^p_{p,e} \  \Big\}^{\frac{1}{p}} \nonumber  \\ 
\overline{\ ^e \mathcal{K}_q^p} (\epsilon) \ & = & \ 
\Big{\{} 
\parallel  r^{(\frac{3}{2} - \frac{3}{p} + q )} 
\ \nlap^q \epsilon \parallel^p_{p, e}  
\ + \ 
\sum_{i+j = q, \ j \geq 1}  \parallel 
r^{ (2 - \frac{2}{p} + i ) } \ (\tau_-)^{1 - \frac{2}{p} + j - 1}  \ 
\nlap^i \nlap^j_N \epsilon
\parallel^p_{p,e} \  \Big\}^{\frac{1}{p}} \nonumber  \\ 
\overline{\ ^e \mathcal{K}_q^p} (\hat{\eta}) \ & = & \ 
\Big{\{} 
\parallel 
r^{(\frac{3}{2} - \frac{3}{p} + q )} \ 
\ \nlap^q \hat{\eta} \parallel^p_{p, e}  
\ + \ 
\sum_{i+j = q, \ j \geq 1}  \parallel r^{ (1 - \frac{2}{p} + i ) } \ (\tau_-)^{1 - \frac{2}{p} + j}  \ 
\nlap^i \nlap^j_N \hat{\eta} 
\parallel^p_{p,e} \  \Big\}^{\frac{1}{p}}   \nonumber  \\ 
& & 
\label{mainthmnormskcopsol*3} \\ 
\nonumber 
\eea
Then we set 
\be
\overline{\ ^e \mathcal{K}_q^p} \ = \ 
\overline{\ ^e \mathcal{K}_q^p} (\delta) \ + \ 
\overline{\ ^e \mathcal{K}_q^p} (\epsilon) \ + \ 
\overline{\ ^e \mathcal{K}_q^p} (\hat{\eta})  
\ee
similarly as above. 
For the case $p=2$, we write simply 
\[
\overline{\mathcal{K}_q} \ = \ \overline{\mathcal{K}_q^2} \ . 
\]
Putting this together into the basic norms of the second fundamental form: 
\bea
\overline{\mathcal{K}_{[0]}} \ & = & \ \overline{\mathcal{K}_0} \\ 
\overline{\mathcal{K}_{[1]}} \ & = & \ \overline{\mathcal{K}_{[0]}} \ + \ 
\overline{\mathcal{K}_1}  \\ 
\overline{\mathcal{K}_{[2]}} \ & = & \ \overline{\mathcal{K}_{[1]}} \ + \ 
\overline{\mathcal{K}_2}  \ .  \\ \nonumber \\ \nonumber 
\eea
\subsubsection{Norms for the Lapse Function $\Phi$} 
\label{norms*lapsefunctphi**subsection*1}
Analogously to the previous cases, we first define 
$\ ^i  \mathcal{L}_q^p$ and $\ ^e  \mathcal{L}_q^p$ to be the interior and exterior 
weighted $L^p$-norms of the $(q+1)$-covariant derivatives of the logarithm $\varphi$ of 
the lapse function $\phi$. Then we also define: 
\be \label{mainthmnormsL**1}
\mathcal{L}_q^p \ = \ \max \big( 
\ ^i  \mathcal{L}_q^p, \ \ ^e  \mathcal{L}_q^p \ \big) \ \ . 
\ee
for $q = 0, 1, 2$. 
Correspondingly, we define 
\be \label{mainthmnormsL**infty*1}
\mathcal{L}_0^{\infty} \ = \ \max \big( 
\ ^i  \mathcal{L}_0^{\infty}, \ \ ^e  \mathcal{L}_0^{\infty} \ \big) \ \ . 
\ee
The interior norms $\ ^ i \mathcal{L}_q^p$ in (\ref{mainthmnormsL**1}) are given by: 
\be \label{compnormsL**3}
 \ ^ i \mathcal{L}_q^p  \ = \ 
r_0^{1 + q - \frac{2}{p}} \parallel D^{q+1} \varphi \parallel_{p, i} \ \ . 
\ee
In order to state the norms for $\ ^ e \mathcal{L}_q^p$, we 
first decompose $\nabla \varphi$ as follows: 
\bea
\philap_A   \ & = & \ \nlap_A \varphi  \label{compnsL**4} \\
\varphi_N \ & = & \ \nabla_N \varphi  \label{compnsL**5} \ \ . 
\eea
That is, we have 
\beas
\philap_A   \ & = & \ \nlap_A \varphi \ = \ \frac{1}{\phi} \ \nlap_A \phi  \\
\varphi_N \ & = & \ \nabla_N \varphi  \ = \ \frac{1}{\phi} \ \nabla_N \phi  \ \ . 
\eeas
Next, we set 
\bea
\ ^e \mathcal{L}_q^p (\philap) \ & = & \ 
\parallel r^{(\frac{3}{2} - \frac{3}{p} + q )} \ \nlap^q \philap \parallel_{p, e}  \nonumber \\ 
\ ^e \mathcal{L}_q^p (\varphi_N) \ & = & \ 
\parallel r^{(\frac{3}{2} - \frac{3}{p} + q )} \ \nlap^q \varphi_N \parallel_{p, e}  \label{compnsL**6} \\ 
\ ^e \mathcal{L}_{q+1}^p (\cDlap_4 \ \philap) \ & = & \ 
\parallel r^{(\frac{5}{2} - \frac{3}{p} + q )} \ \nlap^q \cDlap_4 \ \philap \parallel_{p, e}  \nonumber \\ 
\ ^e \mathcal{L}_{q+1}^p (\cDlap_3 \ \philap) \ & = & \ 
\parallel r^{(\frac{5}{2} - \frac{3}{p} + q )} \ \nlap^q  \cDlap_3 \ \philap \parallel_{p, e}  \nonumber \\ 
\ ^e \mathcal{L}_{q+1}^p (D_4 \ \varphi_N) \ & = & \ 
\parallel r^{(\frac{5}{2} - \frac{3}{p} + q )} \ \nlap^q D_4 \ \varphi_N \parallel_{p, e}  \nonumber \\ 
\ ^e \mathcal{L}_{q+1}^p (D_3 \ \varphi_N) \ & = & \ 
\parallel r^{(\frac{3}{2} - \frac{3}{p} + q )} \tau_-^{\frac{3}{2} - \frac{1}{p} } 
\ \nlap^q D_3 \ \varphi_N \parallel_{p, e}   \label{compnsL**7} \\ 
\ ^e \mathcal{L}_{q+1}^p (\cDlap_S \ \philap) \ & = & \ 
\parallel r^{(\frac{3}{2} - \frac{3}{p} + q )} \ \nlap^q  \philap \parallel_{p, e}  \nonumber \\ 
\ ^e \mathcal{L}_{q+1}^p (\cDlap_S \ \varphi_N) \ & = & \ 
\parallel r^{(\frac{3}{2} - \frac{3}{p} + q )} \ \nlap^q  \varphi_N \parallel_{p, e}  \label{compnsL**8} \\ 
\ ^e \mathcal{L}_{q+2}^p (\cDlap_S \ \cDlap_4 \ \philap) \ & = & \ 
\parallel r^{(\frac{5}{2} - \frac{3}{p} + q )} \ \nlap^q \cDlap_4 \ \philap \parallel_{p, e}  \nonumber \\ 
\ ^e \mathcal{L}_{q+2}^p (\cDlap_S \ \cDlap_3 \ \philap) \ & = & \ 
\parallel r^{(\frac{5}{2} - \frac{3}{p} + q )} \ \nlap^q \cDlap_3 \ \philap \parallel_{p, e}  \nonumber \\ 
\ ^e \mathcal{L}_{q+2}^p (D_S D_4 \ \varphi_N) \ & = & \ 
\parallel r^{(\frac{5}{2} - \frac{3}{p} + q )} \ \nlap^q  D_4 \varphi_N \parallel_{p, e} \nonumber \\ 
\ ^e \mathcal{L}_{q+2}^p (D_S D_3 \ \varphi_N) \ & = & \ 
\parallel r^{(\frac{3}{2} - \frac{3}{p} + q )} \tau_-^{\frac{3}{2} - \frac{1}{p}} \ 
\nlap^q  D_3 \varphi_N \parallel_{p, e}  \label{compnsL**9} \ \ . \\ 
\nonumber 
\eea
Then, we set 
\bea
\ ^e \mathcal{L}_0^p [\philap] 
\ & = & \ 
\ ^e \mathcal{L}_0^p (\philap)  \nonumber \\ 
\ ^e \mathcal{L}_1^p [\philap] 
\ & = & \ 
\ ^e \mathcal{L}_1^p (\philap) \ + \ \ ^e \mathcal{L}_1^p (\cDlap_3 \ \philap) \ + \ 
\ ^e \mathcal{L}_1^p (\cDlap_4 \ \philap)   \nonumber \\ 
\ ^e \mathcal{L}_2^p [\philap] 
\ & = & \ 
\ ^e \mathcal{L}_2^p (\philap) \ + \ \ ^e \mathcal{L}_2^p (\cDlap_3 \ \philap) \ + \ 
\ ^e \mathcal{L}_2^p (\cDlap_4 \ \philap)  \label{compnsL**10} 
\eea
and correspondingly for $\varphi_N$. \\ \\ 
Then, we define for $q = 0, 1, 2$: 
\be \label{compnsL**101} 
\ ^e \mathcal{L}_q^p \ \ = \ \ \ ^e \mathcal{L}_q^p [\philap] \ + \ 
\ ^e \mathcal{L}_q^p [\varphi_N]  \ \ .  
\ee
These norms we use for $p=2$ and we write 
\[
\ ^e \mathcal{L}_q \ \ = \ \ \ ^e \mathcal{L}_q^2 \ \ . 
\]
Finally, we define 
\bea
\ ^i \mathcal{L}_{[0]} \ & = & \ \ ^i \mathcal{L}_0  \nonumber \\ 
\ ^i \mathcal{L}_{[1]} \ & = & \  \ ^i \mathcal{L}_{[0]} \ + \   \ ^i \mathcal{L}_1  \nonumber \\ 
\ ^i \mathcal{L}_{[2]} \ & = & \  \ ^i \mathcal{L}_{[1]} \ + \   \ ^i \mathcal{L}_2  \nonumber \\ 
\ ^e \mathcal{L}_{[0]} \ & = & \ \ ^e \mathcal{L}_0  \nonumber \\ 
\ ^e \mathcal{L}_{[1]} \ & = & \  \ ^e \mathcal{L}_{[0]} \ + \   \ ^e \mathcal{L}_1  \nonumber \\ 
\ ^e \mathcal{L}_{[2]} \ & = & \  \ ^e \mathcal{L}_{[1]} \ + \   \ ^e \mathcal{L}_2   \label{compnsL**102}   
\\ \nonumber  
\eea
and 
\be \label{compnsL**1023}  
\mathcal{L}_{[q]}  \ \ = \ \ \ ^i \mathcal{L}_{[q]} \ + \ \ ^e \mathcal{L}_{[q]}  \ \ . 
\ee 
\subsubsection{Norms for the Hessian of the Optical Function $u$} 
\label{norms*hessoptfctu**subsection*1}
Here, we state the $L^2$-norms for the Hessian $D^2 u$ of the optical function $u$. 
Let $\ ^i  \mathcal{O}_q$, respectively $\ ^e  \mathcal{O}_q$, denote the 
interior, respectively exterior, norms. We set 
\be \label{mainthmnormsO**1}
\mathcal{O}_q \ = \ \max \big( 
\ ^i  \mathcal{O}_q, \ \ ^e  \mathcal{O}_q \ \big) \ \ . 
\ee
for $q = 0, 1, 2$. 
Correspondingly, we define 
\be \label{mainthmnormsO**infty*1}
\mathcal{O}_0^{\infty} \ = \ \max \big( 
\ ^i  \mathcal{O}_0^{\infty}, \ \ ^e \mathcal{O}_0^{\infty} \ \big) \ \ . 
\ee
In the interior region, we express the components of $D^2 u$ with respect to the 
standard null frame in terms of 
$\chi' = a \chi$, $\zeta' = \zeta$, $\omega' = a^{-1} \omega$. 
Let us define in the interior: 
\bea
\ ^i \ \mathcal{O}_q \ (tr \chi' - \overline{tr \chi'}) \ & = & \ 
r_0^{q} \sum_{i+j+k=q} \parallel \nlap^i D_3^j D_4^k \ (tr \chi' - \overline{tr \chi'}) \parallel_{2,i} \nonumber \\ 
\ ^i \ \mathcal{O}_q \ (\hat{\chi}')  \ & = & \ 
r_0^{q} \sum_{i+j+k=q} \parallel \nlap^i \cDlap_3^j \cDlap_4^k \  \hat{\chi}'  
\parallel_{2,i} \nonumber \\ 
\ ^i \ \mathcal{O}_q \ (\zeta')  \ & = & \ 
r_0^{q} \sum_{i+j+k=q} \parallel \nlap^i D_3^j D_4^k \ \zeta'  
\parallel_{2,i} \nonumber \\ 
\ ^i \ \mathcal{O}_q \ (\omega')  \ & = & \ 
r_0^{q} \sum_{i+j+k=q} \parallel \nlap^i D_3^j D_4^k \ \omega'  
\parallel_{2,i}  \ \  .   \label{compnosO***1}  
\eea
Then, we set 
\be 
\ ^i \ \mathcal{O}_q \ = \ 
\ ^i \ \mathcal{O}_q \ (tr \chi' - \overline{tr \chi'}) \ + \ 
\ ^i \ \mathcal{O}_q \ (\hat{\chi}') \ + \ 
\ ^i \ \mathcal{O}_q \ (\zeta') \ + \ 
\ ^i \ \mathcal{O}_q \ (\omega')  \ \ .  \label{compnosO***2}  
\ee
In the exterior region, we work with the $l$-pair and the null frame related to it. 
The components of the Hessian of $u$ in the exterior region with respect to the 
$l$-null frame, that is $tr \chi, \hat{\chi}, \zeta, \omega$, behave differently. 
Let us also introduce 
\bea
tr \chi_4 \ & = & \ 
D_4 tr \chi \ + \ \frac{1}{2} \ tr \chi \ tr \chi  \nonumber \\ 
tr \chi_3 \ & = & \ 
D_3 tr \chi \ + \ \frac{1}{2} \ tr \underline{\chi} \ tr \chi  \nonumber \\ 
\hat{\chi}_4 \ & = & \ 
\cDlap_4 \ \hat{\chi} \ + \ tr \chi \ \hat{\chi} \nonumber \\ 
\hat{\chi}_3 \ & = & \ 
\cDlap_3 \ \hat{\chi} \ + \ tr \underline{\chi} \ \hat{\chi}  \nonumber \\ 
\zeta_4 \ & = & \ 
\cDlap_4 \ \zeta \ + \ \frac{1}{2} \  tr \chi \ \zeta \nonumber \\ 
\zeta_3 \ & = & \ 
\cDlap_3 \ \zeta \ + \  \frac{1}{2} \ tr \underline{\chi} \ \zeta  \nonumber \\  
\omega_4 \ & = & \ 
D_4 \omega \nonumber \\  
\omega_3 \ & = & \ 
D_3 \omega  \ \ .   \label{compnosO***3}  
\eea
Then we define
\bea
\ ^e \ \mathcal{O}_q \ (tr \chi - \overline{tr \chi}) \ & = & \ 
\parallel r^{q} \ \nlap^q (tr \chi - \overline{tr \chi}) \parallel_{2, e} \nonumber \\ 
\ ^e \ \mathcal{O}_q \ (\hat{\chi}) \ & = & \ 
\parallel r^{q} \ \nlap^q \hat{\chi} \parallel_{2, e} \nonumber \\  
\ ^e \ \mathcal{O}_q \ (\zeta) \ & = & \ 
\parallel r^{q} \ \nlap^q \zeta \parallel_{2, e} \nonumber \\  
\ ^e \ \mathcal{O}_q \ (\omega) \ & = & \ 
\parallel r^{- \frac{1}{2} + q}  \tau_-^{\frac{1}{2}} \ \nlap^q \omega \parallel_{2, e} \label{compnosO***4} \\  
\ ^e \ \mathcal{O}_q \ (\chi) \ & = & \ 
\max \big{\{} \ ^e \ \mathcal{O}_q \ (tr \chi - \overline{tr \chi}), \ ^e \ \mathcal{O}_q \ (\hat{\chi})  \big{\}} 
\nonumber \\ 
\ ^e \ \mathcal{O}_{q+1} \ (\hat{\chi}_4) \ & = & \ 
\parallel r^{1+q} \ \nlap^q \hat{\chi}_4 \parallel_{2, e}  \nonumber \\  
\ ^e \ \mathcal{O}_{q+1} \ (\hat{\chi}_3) \ & = & \ 
\parallel r^{1+q} \ \nlap^q \hat{\chi}_3 \parallel_{2, e}  \nonumber \\  
\ ^e \ \mathcal{O}_{q+1} \ (tr \chi_4) \ & = & \ 
\parallel r^{1+q} \ \nlap^q tr \chi_4 \parallel_{2, e}  \nonumber \\  
\ ^e \ \mathcal{O}_{q+1} \ (tr \chi_3) \ & = & \ 
\parallel r^{q} \tau_- \ \nlap^q tr \chi_3 \parallel_{2, e}  \nonumber \\  
\ ^e \ \mathcal{O}_{q+1} \ (\zeta_4) \ & = & \ 
\parallel r^{1+q} \ \nlap^q \zeta_4 \parallel_{2, e}  \nonumber \\  
\ ^e \ \mathcal{O}_{q+1} \ (\zeta_3) \ & = & \ 
\parallel r^{1+q} \ \nlap^q \zeta_3 \parallel_{2, e}  \nonumber \\  
\ ^e \ \mathcal{O}_{q+1} \ (\omega_4) \ & = & \ 
\parallel r^{\frac{1}{2} +q} \tau_-^{\frac{1}{2}} \ \nlap^q \omega_4 \parallel_{2, e}  \nonumber \\  
\ ^e \ \mathcal{O}_{q+1} \ (\omega_3) \ & = & \ 
\parallel r^{- \frac{1}{2} +q} \tau_-^{\frac{3}{2}} \ \nlap^q \omega_3 \parallel_{2, e}  \ \ .  \label{compnosO***5}
\eea
Next, we set 
\bea
\ ^e \ \mathcal{O}_0 [tr \chi] \ & = & \ 
\ ^e \ \mathcal{O}_0 (tr \chi) \nonumber \\ 
\ ^e \ \mathcal{O}_1 [tr \chi] \ & = & \ 
\ ^e \ \mathcal{O}_1 (tr \chi) \ + \ 
\ ^e \ \mathcal{O}_1 (tr \chi_3) \ + \ \ ^e \ \mathcal{O}_1 (tr \chi_4)  \nonumber \\ 
\ ^e \ \mathcal{O}_2 [tr \chi] \ & = & \ 
\ ^e \ \mathcal{O}_2 (tr \chi) \ + \ 
\ ^e \ \mathcal{O}_2 (tr \chi_3) \ + \ \ ^e \ \mathcal{O}_2 (tr \chi_4)  \ \ .   \label{compnosO***6}
\eea
Correspondingly, we proceed for $\hat{\chi}, \zeta, \omega$. \\ \\ 
Now, for $q = 0, 1, 2$ one sets 
\be \label{compnosO***7}
\ ^e \ \mathcal{O}_q \ = \ \ ^e \ \mathcal{O}_q^2 \ = \ 
\ ^e \ \mathcal{O}_q [tr \chi] \ + \ 
\ ^e \ \mathcal{O}_q [\hat{\chi}] \ + \  
\ ^e \ \mathcal{O}_q [\zeta] \ + \  
\ ^e \ \mathcal{O}_q [\omega]  \ \ . 
\ee
Then we set 
\bea
\ ^i \ \mathcal{O}_{[0]}^{\infty} \ & = & \ 
\ ^i \ \mathcal{O}_{0}^{\infty} \ + \ 
r_0^{\frac{1}{2}} \ \sup_{I} \mid \overline{tr \chi} - \frac{2}{r} \mid \ + \ 
\sup_{I} \mid a \ - \ 1 \mid \nonumber \\ 
\ ^e \ \mathcal{O}_{[0]}^{\infty} \ & = & \ 
\ ^e \ \mathcal{O}_{0}^{\infty} \ + \ 
\sup_{r \geq \frac{r_0}{2}} r^{\frac{1}{2}} \mid \overline{tr \chi} - \frac{2}{r} \mid \ + \ 
\sup_{r \geq \frac{r_0}{2}}  \mid a \ - \ 1 \mid  \ \ .  \label{compnosO***8} 
\eea
Finally, we define 
\bea
\ ^i \ \mathcal{O}_{[0]} \ & = & \ 
\ ^i \ \mathcal{O}_{0} \ + \ \ ^i \ \mathcal{O}_{[0]}^{\infty} \nonumber \\ 
\ ^i \ \mathcal{O}_{[1]} \ & = & \ 
\ ^i \ \mathcal{O}_{[0]} 
\ + \ \ ^i \ \mathcal{O}_{1}  \nonumber \\ 
\ ^i \ \mathcal{O}_{[2]} \ & = & \ 
\ ^i \ \mathcal{O}_{[1]} 
\ + \ \ ^i \ \mathcal{O}_{2}  \nonumber \\ 
\ ^e \ \mathcal{O}_{[0]} \ & = & \ 
\ ^e \ \mathcal{O}_{0} \ + \ \ ^e \ \mathcal{O}_{[0]}^{\infty} \nonumber \\ 
\ ^e \ \mathcal{O}_{[1]} \ & = & \ 
\ ^e \ \mathcal{O}_{[0]} 
\ + \ \ ^e \ \mathcal{O}_{1}  \nonumber \\ 
\ ^e \ \mathcal{O}_{[2]} \ & = & \ 
\ ^e \ \mathcal{O}_{[1]} 
\ + \ \ ^e \ \mathcal{O}_{2}  \label{compnosO***zu*1}   
\eea
and 
\be  \label{compnosO***zu*2}  
\mathcal{O}_{[q]} \ \ = \ \ \ ^i \ \mathcal{O}_{[q]}  \ + \  \ ^e \ \mathcal{O}_{[q]}  \ \ . 
\ee 
\[\]
\subsection{Bootstrap Argument}
\label{norms*BA**subsection*1}
The method of bootstrapping constitutes the core of the proof of the main theorem. 
Several steps are needed to close the bootstrap loop. \\ \\ 
We define $\mathcal{S}$ to be the set of all $t \geq 0$ such that there exists a spacetime slab 
$\bigcup_{t' \in [0,t]} H_{t'}$ endowed with a canonical optical function with respect to which 
the following 
\index{bootstrap assumptions!}
\begin{itshape}bootstrap assumptions \end{itshape} hold: 
\begin{itemize}
\item {\bf BA0:} For all 
$t' \in [0,t]$, assume that 
\be \label{BA0}
\frac{1}{2} \ (1 \ + \ t') \ \leq \ r_0 (t') \ \leq \ \frac{3}{2} \ (1 \ + \ t') \ . 
\ee
\item {\bf BA1:} For all 
$t' \in [0,t]$, assume that 
\be \label{BA1}
\ ^e\mathcal{O}_0^{\infty}, \ \mathcal{K}_0^{\infty}, \ \mathcal{L}_0^{\infty} \ \ 
\leq \ \ \epsilon_0 \ . 
\ee
\item {\bf BA2:} For all 
$t' \in [0,t]$, assume that 
\be \label{BA2}
\mathcal{R}_{[1]}, \ 
\ ^e\mathcal{O}_{[2]}, \ 
\mathcal{K}_{[2]}, \ \mathcal{L}_{[2]} \ \ 
\leq \ \ \epsilon_0 \ . 
\ee
\end{itemize}
Additional Assumption on the last slice $H_t$: 
\be \label{addass**la1}
\sup_{S_u} \big{\{} r^{\frac{3}{2}} \mid \nlap \log a \big{\}} \ \ \leq \ \ \epsilon_0 \ . 
\ee
First, we show that the set $\mathcal{S}$ is not empty. 
Of course, the bootstrap assumptions hold at $t=0$.  
Afterwards, 
we define the supremum of $\mathcal{S}$ as $t_*$. If $t_* = \infty$, then 
the global existence is proved. We use a continuity argument 
to derive that indeed $t_* = \infty$. This goes by contradiction, assuming that 
$t_* < \infty$. 
Then, it is $t_* \in \mathcal{S}$, 
and the inequalities must be saturated at $t_*$. It will be shown that this cannot happen. This yields that 
$t_* = \infty$. \\ \\ 
{\bf Step 1:} 
We show that the set $\mathcal{S}$ is not empty. That is, it contains at least $t=0$. 
In view of the local existence theorem we can construct the past slab 
$\bigcup_{t' \in [-1,0]} H_{t'}$ and the initial cone $C_0$ with vertex at a point 
on $H_{-1}$. Then an exterior optical function $u$ on $t=0$ is constructed by solving 
the inverse lapse problem, starting on the 2-surface $S_{0,0} = C_0 \cap H_0$, 
according to the procedure explained in Step 4 for `The Last Slice'.  
We proceed analogously to Step 4 for 
a background radial function $u'$ on $H_{0}$. 
One then shows that the norms 
$\mathcal{R}_{[1]}, \mathcal{K}_{[2]}, \mathcal{L}_{[2]},  \ ^e\mathcal{O}_{[2]}$ 
as well as 
$\ ^e\mathcal{O}_0^{\infty},  \mathcal{K}_0^{\infty},  \mathcal{L}_0^{\infty}$ 
can be made arbitrarily small, that is the assumptions {\bf BA1} and {\bf BA2} are fulfilled. 
The additional assumption (\ref{addass**la1}) follows from the main result in Step 4.  \\ \\ 
{\bf Step 2:} 
Let $t_* = \sup \mathcal{S}$. If $t_* = \infty$, then the global existence is proved. 
Now, we assume 
$t_* < \infty$. Then, it is $t_* \in \mathcal{S}$. 
The hypersurface $H_{t_*}$ is called the `last slice' of the $t$-foliation of the bootstrap argument. 
One extends the exterior optical function $u$ of the spacetime slab $\bigcup_{t \in [0, t_*]} H_{t}$ to 
the interior region, for which one derives the following inequalities: 
\bea
\ ^i \mathcal{O}_{[2]} \ & \leq & \ 
c \ \big( \mathcal{R}_{[1]} \ + \ 
\mathcal{K}_{[2]} \ + \  \mathcal{L}_{[2]}  \big) \nonumber \\ 
\ ^i \mathcal{O}_{0}^{\infty} \ & \leq & \ 
c \ \big( 
\mathcal{R}_{[1]} \ + \ 
\mathcal{K}_0^{\infty} \ + \ \mathcal{L}_0^{\infty} 
 \big) \ \ . 
\label{BAstep2ass*s1}
\eea
Then one obtains  
\be \label{BAstep2res***1}
\ ^i \mathcal{O}_{[2]} \ , \  \ ^i \mathcal{O}_{0}^{\infty} \ \ \  \leq  \ \ \ 
c \ \epsilon_0 \ \ . 
\ee
The global function $u$ is obtained by the matching of two optical functions, defined 
in the exterior and in the interior region respectively. The exterior optical function is crucial, as it describes 
the structure of null infinity. And one constructs it by solving the Eikonal equation with initial conditions on the last slice. 
For the interior optical function one prescribes initial conditions on a `central line' given by an integral curve of the vectorfield $T$. 
The quantities related to the foliation given in the exterior region, are more subtle to estimate than the ones 
in the interior. Thus, proving the exterior estimates takes the largest part of the work. 
Further information is given in \cite{lydia1}, \cite{lydia2} and \cite{sta}. \\ \\ \\ 
{\bf Step 3:} 
This is the core part of the proof. 
It splits into three sections, namely, obtaining estimates for 
\begin{itemize}
\item[{\bf a)}] $\mathcal{R}_{[1]}$, 
\item[{\bf b)}] $\mathcal{K}_{[2]}, \mathcal{L}_{[2]}$, respectively, $\mathcal{K}_0^{\infty}, \mathcal{L}_0^{\infty}$, 
\item[{\bf c)}] $\ ^e \mathcal{O}_{[2]}$, respectively, $\ ^e \mathcal{O}_{0}^{\infty}$. 
\end{itemize}
Using step 2 and the bootstrap assumptions {\bf BA0}, {\bf BA1}, {\bf BA2}, we show that 
the size of the norms 
$\ ^e\mathcal{O}_0^{\infty},  \mathcal{K}_0^{\infty},  \mathcal{L}_0^{\infty}$ 
and
$\mathcal{R}_{[1]}, \mathcal{K}_{[2]}, \mathcal{L}_{[2]},  \ ^e\mathcal{O}_{[2]}$ 
cannot exceed a constant multiple of the respective size of the data at $t=0$. 
Therefore, one can choose $\epsilon$ and $\epsilon_0$ sufficiently small such that 
\bea
\ ^e\mathcal{O}_0^{\infty}, \  \mathcal{K}_0^{\infty}, \ \mathcal{L}_0^{\infty} \ \ 
& \leq & \ \ \frac{1}{2} \  \epsilon_0 \  \nonumber \\ 
\ ^e\mathcal{O}_{[2]}, \ \mathcal{R}_{[1]}, \ 
\mathcal{K}_{[2]}, \ \mathcal{L}_{[2]} \ \ 
& \leq & \ \ \frac{1}{2} \  \epsilon_0 \ . 
\eea
This is achieved as follows: \\ \\ 
{\bf a)} 
In this part, 
we use the bootstrap assumptions {\bf BA0}, {\bf BA1}, {\bf BA2} and inequality (\ref{BAstep2res***1}) 
to check all the assumptions of the comparison theorem, which in item 2. of the main steps of the proof of our 
main result laid out at the beginning of section 5, 
allows us to estimate 
the curvature by $Q_1(W)$, 
as well as of the main theorem concerning the error estimates, where we estimate the 
three terms of $Q_1(W)$, as given in (\ref{intQ1W12})-(\ref{intQ1W123}).  \\ \\ 
According to the main theorem for the error estimates, that is from item 1. of the main steps of the proof of our 
main result laid out at the beginning of section 5, 
we have 
\be \label{BAres***Q1}
Q_1* \ + \ Q_0* \ \leq \ c \ ( Q_1 (0) \ + \ Q_0 (0) ) \ . 
\ee
Thus by the comaprison theorem of item 2. at the beginning of section 5, 
we conclude that, for all $t \in [0,t_*]$, 
\be \label{BAres***R1}
\mathcal{R}_{[1]} (t) \ \leq \ c \ \mathcal{R}_{[1]} (0) \ . 
\ee
We recall that in our work, we do not need 
rotational vectorfields at all to derive the estimates for 
$\mathcal{R}_{[1]} (t)$. Contrary to \cite{sta}, where rotational vectorfields 
were used in a crucial way, 
we only work with 
the vectorfields $T$, $S$ and $\bar{K}$ in conjunction with the Bianchi equations 
to deduce the required estimates for $\mathcal{R}_{[1]} (t)$. 
That is, we define the quantities $Q_0$ and $Q_1$ with help of the vectorfields 
$T$, $S$ and $\bar{K}$ as given also above. 
Then, we estimate $\mathcal{R}_{[1]} (t)$ in terms of 
$Q_0$ and $Q_1$ by the comparison argument. \\ \\ 
From (\ref{BAres***R1}) and from the fact that we can bound 
$ \mathcal{R}_{[1]} (0)$ by $c \cdot \epsilon$, 
we decuce 
\be \label{BAres***R1*3}
\mathcal{R}_{[1]} (t) \ \leq \ c \ \epsilon \ \ . 
\ee
Choosing $\epsilon$ sufficiently small, yields 
\be \label{BAres***R1*4}
\mathcal{R}_{[1]} (t) \ \leq \ \frac{1}{2} \ \epsilon_0 \ \ . 
\ee
\[\]
{\bf b)} 
We show that the bootstrap assumptions 
{\bf BA0}, {\bf BA1}, {\bf BA2} and the inequality (\ref{BAstep2res***1}) imply the following: 
\[
\mathcal{K}_{[2]}, \ \mathcal{L}_{[2]} \ \leq \ c \ \mathcal{R}_{[1]} \ \ . 
\]
Then, by the Sobolev inequalities, we deduce that 
\[
\mathcal{K}_0^{\infty}, \ \mathcal{L}_0^{\infty} \ \leq \ c \ \mathcal{R}_{[1]} \ , 
\]
and therefore, choosing $\epsilon$ sufficiently small, we conclude 
\bea
\mathcal{K}_0^{\infty}, \ \mathcal{L}_0^{\infty} \ & \leq & \ \frac{1}{2} \ \epsilon_0 \\ 
\mathcal{K}_{[2]}, \ \mathcal{L}_{[2]} \ & \leq & \ \frac{1}{2} \ \epsilon_0 \ . 
\eea
{\bf c)} 
Here, 
we show that the bootstrap assumptions 
{\bf BA0}, {\bf BA1}, {\bf BA2} imply the following: 
\beas
\ ^e \mathcal{O}_0^{\infty} \ & \leq & \ c \ \epsilon \\ 
\ ^e \mathcal{O}_{[2]}  \ & \leq & \ c \ \epsilon \ . 
\eeas
Therefore, if $\epsilon$ is sufficiently small, this yields 
\bea
\ ^e \mathcal{O}_0^{\infty} \ & \leq & \ \frac{1}{2} \ \epsilon_0 \\ 
\ ^e \mathcal{O}_{[2]}  \ & \leq & \ \frac{1}{2} \ \epsilon_0 \ . 
\eea
\[\]
{\bf Step 4:} 
This step is to be considered together with the previous one. However, 
as it is a crucial point within the whole procedure of the bootstrap argument, 
we formulate it separately. \\ \\ 
We show that we can extend our spacetime beyond the time $t_*$. In particular, we use the 
result of the previous step together with the local existence theorem, 
with initial data at $t_*$, to extend the spacetime up from $t_*$ to $t_* + \delta$. 
Also, the optical function $u'$ of the spacetime slab 
$\bigcup_{t \in [0, t_*]} H_t$ 
is extended 
by continuing the null geodesic generators of the hypersurfaces $C_{u'}$ into the future up to 
$t_* + \delta$. 
We choose $\delta$ to be sufficiently small, such that the size of the norms 
$\mathcal{R'}_{[1]}, \mathcal{K'}_{[2]},  \mathcal{L'}_{[2]}, \ ^e \mathcal{O'}_{[2]}$ and 
$\mathcal{K'}_0^{\infty}, \mathcal{L'}_0^{\infty}, \ ^e \mathcal{O'}_0^{\infty}$ 
remains strictly smaller than $\epsilon_0$. 
Moreover, 
$\sup_{S'_{u'}} \big{\{} r'^{\frac{3}{2}} \mid \nlap' \log a' \big{\}}$ in 
(\ref{addass**la1}) is strictly smaller than $\epsilon_0$. \\ \\ 
Now, we start with $H_{t_* + \delta}$ as last slice. 
The cut 
$S_{t_* + \delta, \ 0} = H_{t_* + \delta} \cap C_0$ 
of this last slice with $C_0$ is the initial 2-surface, 
from which we start, 
to solve 
the appropriate equation of motion of surfaces 
and construct a new 
optical function $u$ on $H_{t_* + \delta}$.  
We recall that the function 
$u'$ gives the background foliation here. As the 
level surfaces of $u$ in $H_{t_* + \delta}$ do not exist yet, 
but have to be constructed, 
we use the bootstrap assumptions on the quantities of the background foliation 
and the comparison between the two foliations induced by $u'$ and $u$ 
to control the curvature and geometric quantities of the foliation by $u$. 
Thus, using a bootstrap argument, we construct the new optical function $u$ on $H_{t_* + \delta}$ 
by solving 
an equation of motion of surfaces on 
$H_{t_* + \delta}$ starting from 
$S_{t_* + \delta, \ 0}$. 
This is then extended to the past. \\ \\ 
In view of the continuity properties of the equations, 
we deduce that the new norms 
$\mathcal{R}_{[1]}$, $\mathcal{K}_{[2]}$,  $\mathcal{L}_{[2]}$, $\ ^e \mathcal{O}_{[2]}$ and 
$\mathcal{K}_0^{\infty}$, $\mathcal{L}_0^{\infty}$, $\ ^e \mathcal{O}_0^{\infty}$ 
can be made arbitrarily close to the previous ones by choosing $\delta$ suitably small. 
One therefore checks that the bootstrap assumptions BA1 and BA2 as well as 
BA0 and inequality (\ref{addass**la1}) 
still hold. 
Thus, we obtain that 
$t_* + \delta \in \mathcal{S}$, which contradicts the assumption that $t_* < \infty$. \\ \\ \\ 
{\bf Step 5:} 
To complete the proof of the main theorem, one shows that the optical function 
$\ ^{(t)} u$ defined on the slab 
$\bigcup_{t' \in [0, t]} H_{t'}$, starting from the last slice $H_{t}$ 
in the exterior 
approaches a global exterior otpical function $u$ as $t \to \infty$. \\ \\ 
\section{Discussion and Outline of the Proof} 
\label{proofoutline*2}
\subsection{General Ideas and Concepts, Curvature and Energy}
There are several important properties of our spacetime which play a crucial role in our proof. 
We are going to discuss them now and also give an outline of the proof of our main theorem (theorem \ref{maintheoremlb2}, respectively of the full version of our main theorem \ref{maintheoremlb*1}). \\ \\ 
In order to obtain the precise estimates, it is necessary to work with appropriate foliations of the spacetime 
$(M, g)$. 
These are the foliation into hypersurfaces $H_t$ given by the time function $t$ and the foliation  
into $u$-null-hypersurfaces $C_u$ by the optical function $u$, as they were introduced earlier.  
The slices $H_t$ of our spacetime are 3-dimensional, complete, Riemannian manifolds, diffeomorphic to $\real^3$ and 
Euclidean at infinity. Each slice carries a structure induced by the level hypersurfaces of the 
optical function $u$. 
Thus, 
the $(t,u)$ foliations of the spacetime $(M,g)$ define a codimension-$2$-foliation by $2$-surfaces 
$S_{t,u} \ = \ H_t \ \cap \ C_u$.  \\ \\ 
The \begin{itshape}asymptotic behaviour \end{itshape} 
of the 
\begin{itshape}curvature tensor \end{itshape} $R$ and the 
\begin{itshape}Hessian \end{itshape} of $t$ and $u$ 
can only be \begin{itshape}fully described \end{itshape} by decomposing them 
into \begin{itshape}components tangent and normal to \end{itshape} $S_{t,u}$. \\ \\ 
One achieves this by introducing  
\begin{itshape}null pairs \end{itshape} 
consisting of two future-directed null vectors $e_4$ and $e_3$ orthogonal to $S_{t,u}$ with 
$e_4$ tangent to $C_u$ and 
\be
\big< e_4,  e_3 \big> \ = \ - \ 2 \ . 
\ee
Note that the null pair $(e_3, e_4)$ is only determined up to a transformation of the form 
\[
(e_3, e_4) \ \mapsto \ (a^{-1} e_3, a e_4) \ , \ \ a > 0 \ . 
\]
It is uniquely determined if we also impose the condition that 
$e_4 - e_3$ is tangential to the $H_t$. \\ \\ 
The null pair is completed with an orthonormal frame $e_1$, $e_2$ on $S_{t,u}$ to form a 
\begin{itshape}null frame. \end{itshape} 
The \begin{itshape}null decomposition \end{itshape} of a tensor relative to a null frame 
$e_4, e_3, e_2, e_1$ is obtained by 
taking \begin{itshape}contractions \end{itshape} with the vectorfields $e_4, e_3$. \\ \\ 
Also define  
\[
\tau_-^2 \ : =  \ 1 \ + \ u^2  \ .
\]
With respect to this frame, we obtain the following 
\begin{itshape} null decomposition of the Riemann curvature tensor of an EV spacetime, \end{itshape} 
where the capital indices take the values $1,2$: 
\bea
R_{A3B3} \ & = & \ \underline{\alpha}_{AB} \label{intnullcurvalphaunderline*1} \\ 
R_{A334} \ & = & \ 2 \ \underline{\beta}_A \\ 
R_{3434} \ & = & \ 4 \ \rho \\ 
\ ^* R_{3434} \ & = & \ 4 \ \sigma \\ 
R_{A434} \ & = & \ 2 \ \beta_A \\ 
R_{A4B4} \ & = & \ \alpha_{AB} 
\eea
with \\ 
\begin{tabular}{lll}
$\alpha$, $\underline{\alpha}$ & : & $S$-tangent, symmetric, traceless tensors \\ 
$\beta$, $\underline{\beta}$ & : &  $S$-tangent $1$-forms \\ 
$\rho$, $\sigma$ & : & scalars \ . \\ 
\end{tabular} 
\\ \\ 
We show, as a part of our main result, that these components are controlled in the sense 
of the main theorem.  
And our estimates yield the decay behaviour: 
\beas
\underline{\alpha} \ & = & \ O \ ( r^{- 1} \ \tau_-^{- \frac{3}{2}}) \\ 
\underline{\beta} \ & = & \ O \ ( r^{- 2} \ \tau_-^{- \frac{1}{2}}) \\ 
\rho , \ \sigma , \ \alpha , \ \beta \ & = & \ o \ (r^{- \frac{5}{2}})  
\eeas
At this point, let us recall that in \cite{sta} the null components have the 
decay properties:
\beas 
\underline{\alpha} \ & = & \ O \ (r^{- 1} \ \tau_-^{- \frac{5}{2}}) \\ 
\underline{\beta} \ & = & \ O \ (r^{- 2} \ \tau_-^{- \frac{3}{2}}) \\ 
\rho \   & = & \ O \ (r^{-3})  \\ 
\sigma  \ & = & \ O \ (r^{-3} \ \tau_-^{- \frac{1}{2}}) \\ 
\alpha , \ \beta \ & = & \ o \ ( r^{- \frac{7}{2}})  
\eeas
The fact that in \cite{lydia1}, \cite{lydia2} only 
one derivative of the curvature (Ricci) in $H$ is controlled, 
means that the 
curvature is 
not pointwise bounded. What one has is only 
the following, where 
$Ric$ includes corresponding weights according to (\ref{globalsalbQ}):
\[ 
Ric \ \in \ W^{1,2} (H) \ . 
\]
The trace lemma gives 
for the Gauss curvature $K$ in the leaves of the $u$-foliation $S$: 
\[
K \ \in \ L^4(S) \ . 
\]
Whereas in \cite{sta}, the authors control two derivatives of the curvature in $L^2(H)$, giving 
$L^{\infty}(H)$ bounds: 
$Ric$ including weights as in \cite{sta}: 
\[
Ric \ \in \ L^{\infty} (H)  \ . 
\]
This yields for the Gauss curvature $K$ in the surfaces $S$ that 
$K \ \in \ L^{\infty}(S)$. 
In our case, we also control two derivatives of the 
second fundamental form $k$. Therefore, by Sobolev inequalities, it is 
\[
k \ \in \ L^{\infty} (H)  \ . 
\]
Then, also in the surfaces $S$, the second fundamental form $k$ lies in $L^{\infty} (S)$. \\ \\ \\
Working with this approach, there are two main difficulties to be discussed. 
They have been stated and solved by Christodoulou and Klainerman in \cite{sta}. 
As these concepts are also crucial in our work, let us now say what they are. 
However, employing these concepts in our setting requires fundamentally new ideas 
in our proofs. It shall be explained below. 
Now, the said difficulties are: \\ \\ 
1) `Energy estimates'.  \\ \\ 
2) A general spacetime has no symmetries. Thus the conformal isometry group is trivial. 
Hence, the vectorfields needed to construct conserved quantities do not exist. \\ \\ 
1) As the goal now is to find estimates for the spacetime curvature to give control on regularity, 
one way to attack the problem could be to focus on the definition of the 
energy-momentum tensor appropriate to a geometric Lagrangian, namely considering the variation of 
the action $\mathcal{A}$ with respect to the underlying metric. 
Generally, for a domain $D$ with compact closure in $M$ and Lagrangian $L$ 
the action $\mathcal{A}$ is defined as: 
\be \label{intactionA1}
\mathcal{A} [D] \ = \ \int_{D} L \ d \mu_g \ \ . 
\ee
Variations supported in $D$ of the action, with respect to the underlying metric, 
yield the energy-momentum tensor as follows: 
\be \label{intactionAvar1}
\dot{\mathcal{A}} [D] \ = \ - \ \frac{1}{2} \ \int_D T^{\mu \nu} \ \dot{g}_{\mu \nu} \ d \mu_g \ \ . 
\ee
But this approach would not work here (or in \cite{sta}), because the variation (\ref{intactionAvar1}) 
vanishes for the gravitational Lagrangian: $L=-\frac{1}{4}R d \mu_{g}$, that is for the 
Einstein-Hilbert action: 
\be
\label{intactionA2}
\mathcal{A} [D] \ = \ - \ \frac{1}{4}  \ \int_{D} R \ d \mu_g \ \ . 
\ee
This vanishing is stated in the Euler-Lagrange equations for gravitation, that is in the 
EV equations. \\ \\ 
An alternative, one could think, could be Noether's theorem after subtracting an appropriate 
divergence relative to a background metric. But the energy could give control on the solutions 
only after the isoperimetric constant is controlled. Therefore, the energy alone could not help to prove 
regularity. For a discussion of this problem, see also \cite{chrdlbmathpgrt}. \\ \\ 
However, the right way to resolve the first difficulty is the following: 
Consider the Bianchi identities: 
\be
D_{[ \alpha} R_{\beta \gamma ] \delta \epsilon} \ = \ 0 \  
\ee
as differential equations for the curvature and the Einstein equations: $R_{\mu \nu} = 0$, as 
algebraic conditions on the curvature. 
At this point, 
breaking the connection between the metric and the curvature, 
one introduces the \begin{itshape}Weyl tensorfield \end{itshape} $W$ of a given spacetime $(M, g)$, which 
has all the symmetry properties of the curvature tensor and 
in addition is traceless 
\be
g^{\alpha \beta} \ W_{\alpha \mu \beta \nu} \ \ = \ \ 0 \ \ , 
\ee
which is the analogue of the Einstein equations, 
and satisfies 
the Bianchi equations,  
\be
D_{[\epsilon} W_{\alpha \beta ] \gamma \delta} \ = \ 0 \ . 
\ee
We remark that the Bianchi equations are linear. \\ \\ 
Note that the Riemann curvature tensor has 20 independent components, whereas the 
conformal curvature and Ricci tensors have 10 components each.  \\ \\ 
The Bel-Robinson tensor $Q$ is defined out of the Weyl tensor $W$ and 
given above in (\ref{intQBelRob1}). 
The main properties of $Q$ are stated after its definition (\ref{intQBelRob1}). 
This quantity $Q$ can 
then be thought of as the 'energy-momentum tensor' in our setting.  \\ \\ 
The Bel-Robinson tensor $Q$, in fact, plays the same role for solutions of the Bianchi equations 
as the energy-momentum tensor of an electric-magnetic field plays for the solutions of 
the Maxwell equations. \\ \\ 
Assume to be given three vectorfields $X, Y, Z$, each of which generating a 
$1$-parameter group of conformal isometries of the spacetime $(M, g)$. 
Then the $1$-form 
\[
P \ = \ - \ Q(\cdot, X, Y, Z) 
\]
is divergence-free. It follows thus that the integral on a Cauchy hypersurface $H$ 
\[
\int_H \ ^* P 
\]
is conserved (and is positive definite, if all of the vectorfields $X, Y, Z$ are 
timelike future-directed), 
recalling that $ \ ^* P$ is the dual $3$-form: 
$\ ^* P_{\mu \alpha \beta} \ = \ P^{\nu} \epsilon_{\nu \mu \alpha \beta}$.  \\ \\ 
To investigate the Einstein equations, being hyperbolic and nonlinear, we use 
energy estimates of the type in \cite{sta}. 
Aiming at global results, the classical energy estimates could not be used, as they 
are only applied for solutions being local in time. 
Instead, 
we introduce energies $Q_0(t)$ and $Q_1(t)$ (see definition above), 
being integrals over $H_t$ involving the Bel-Robinson tensor $Q$ of the spacetime curvature 
$W$ and of the Lie derivatives of $W$, 
which serve to estimate the curvature components by a comparison argument. 
This is one of the core parts of our work, and it 
is different from the work of D. Christodoulou and S. Klainerman in a fundamental way, which 
will be explained below. \\ \\ 
The quantities $Q_0(t)$ and $Q_1(t)$ themselves are estimated by a continuity argument 
to be bounded by a multiple of the initial value $Q_1(0)$. 
More precisely, we show the error terms, that are generated while estimating 
the growth of $Q_0$ and $Q_1$, 
to be controlled. In this procedure it is important to assess the structure of these 
nonlinear terms. It turns out that the most troublesome terms cancel by identities that are 
consequences of the covariance and algebraic properties of the Einstein equations. \\ \\ 
As a major result emerges the fact that the estimates for 
the most delicate of these error terms are borderline. This means, that any 
further relaxation of the assumptions would lead to divergence and the argument would 
not close anymore. This needs further explanation. 
Contrary to many problems in analysis, where the 
principal terms, that is the terms containing the highest derivatives,  
are the most sensitive ones to estimate, 
whereas the non-principal terms (containing less or no derivatives)
are usually easier to handle, here 
the most difficult terms to be estimated are of 
higher order with respect to asymptotic behaviour (that is they have less decay), 
but they are non-principal 
from the point of view of differentiability. 
On the other hand the expressions, which are principal with respect to derivatives 
behave better asymptotically, and therefore can be controlled easier. 
Thus, by `borderline' we always mean borderline 
from the point of view of decay (asymptotic behaviour). 
It is an essential difference between the situation investigated by 
D. Christodoulou and S. Klainerman in \cite{sta} and ours that 
their worst terms still being of lower order in asymptotic behaviour than ours, the 
borderline case does not appear, whereas in our setting the estimates for the 
highest order terms in view of asymptotic behaviour are really borderline. \\ \\ \\
2) The second difficulty is, that a general spacetime has no symmetries, that is the 
conformal isometry group is trivial. Then one could not construct integral conserved quantities by using 
vectorfields in conjunction with energy-momentum tensors. \\ \\ 
The solution is as follows: 
As a spacetime arising from arbitrary asymptotically flat initial data is itself supposed to be 
asymptotically flat at spacelike infinity in general, and also, with corresponding smallness 
assumptions of the initial data, as the time tends to infinity, one could expect the spacetime 
to approach Minkowski spacetime. Now, Minkowski spacetime has a 
large conformal isometry group. The idea is, to use part of it in the following way. 
One defines in the limit an action of a subgroup. 
Next, one extends this action backwards in time up to the initial hypersurface in a manner as to 
obtain an action of the said subgroup globally. 
This has to be done in a way such that the deviation from conformal isometry is globally small and 
goes to zero at infinity sufficiently rapidly. It is described by the 
circumstance that the trace-free part $\ ^{(X)} \hat{\pi}$ of 
the deformation tensor $\ ^{(X)} \pi : = \mathcal{L}_X g$ 
of the generating vectorfield $X$ 
is globally small and approaches zero sufficiently fast at infinity. 
In order to derive a complete system of estimates, 
we define the action of the subgroup of the conformal group of Minkowski spacetime 
corresponding to 
the time translations, the scaling and the inverted time translations. 
We recall that, contrary to the work \cite{sta} of S. Christodoulou and S. Klainerman, 
our proof does not involve any rotational vectorfields. \\ \\ 
The action of the group of time translations is the easisest to define. 
Having chosen a canonical maximal time function $t$, the corresponding 
time translation vectorfield $T$ generates the action, taking the 
maximal hypersurfaces into each other, as described above. 
This and the optical function $u$, from which 
the action of the other groups are defined, are introduced above, where 
we discuss the $(t,u)$-foliation of the spacetime.  \\ \\ 
Let us define the vectorfields $S$ for the scaling and $K$ for the inverted time translation. 
First, we introduce the function $\underline{u}$ to be 
\be \label{intuunderl}
\underline{u} \ = \ u \ + \ 2 \ r \ \ .
\ee
The time translation vectorfield $T$ has already been defined. We only remark here, that it can be 
written as in the subsequent formula. 
Let $L$ and $\underline{L}$ be the outgoing, respectively incoming, null normals to the 
surface $S_{t,u}$ given by (\ref{intSHC1}), 
for which the component along $T$ is equal to $T$. Also, 
the integral curves of $L$ are the null geodesic generators of the null hypersurfaces $C_u$ parametrized 
by $t$. \\ \\ 
Then $T$ is expressed as 
\be \label{intT12}
T \ = \ \frac{1}{2} \ \big( L \ + \ \underline{L} \big)  \ \ . 
\ee
The generator $S$ of scalings is defined to be: 
\be \label{intS12}
S \ = \ \frac{1}{2} \ \big( \underline{u} \ L \ + \ u \ \underline{L}  \big) \ \ . 
\ee
And the generator $K$ of inverted time translations is defined as: 
\be \label{intK12}
K \ = \ \frac{1}{2} \ \big( \underline{u}^2 \ L \ + \ u^2 \ \underline{L}  \big) \ \ . 
\ee
Then the vectorfield $\bar{K} = K + T$ reads as: 
\be \label{intKbar12}
\bar{K} \ = \ \frac{1}{2} \ \big( \tau_+^2 \ L \ + \ \tau_-^2 \ \underline{L}  \big) \ \ . 
\ee
These vectorfields are used to construct 
quantities whose growth can be controlled in terms of the quantities themselves. 
This procedure is in the spirit of the theorem of Noether. 
We observe that it is crucial to work with a characteristic foliation of the spacetime 
in order to obtain the required quasi-conformal isometries.  \\ \\ 
In the subsequent paragraphs, we are going to outline the method, how 
the energies $Q_0$ and $Q_1$ are constructed and estimated. \\ \\ 
Before, let us say a few words about the following. 
As pointed out in the main step 3 above, we have to estimate the geometric quantities by 
curvature assumptions and using mainly elliptic estimates on the surfaces $S_{t,u}$ and evolution 
equations in $H_t$ and $C_u$. 
The elliptic tools are used in many situations. In particular, they are crucial in deriving 
inequalities for the components of the second fundamental form. 
Note that in the estimates we have to make a difference between 
angular and normal components and derivatives relative to the radial foliation. 
This is a consequence of the 
decay behaviour of the spacetime curvature (in the wave zone). 
Following the notation of \cite{sta}, we call such estimates degenerate, and the usual type non-degenerate. 
At several points, we work with $L^p$ estimates on the surfaces $S$. 
In order to obtain them, we need the uniformization theorem so that we can use 
the Calderon-Zygmund theory for the corresponding Hodge systems on the standard sphere. 
We prove the uniformization theorem in \cite{lydia2} and in \cite{lydia1} 
in chapter 9, theorem 13  
for our setting, where the Gauss curvature $K$ is in 
$L^4(S)$.  \\ \\ 
As a central part of the proof, we state and prove the 
comparison theorem from the main step 2 above. 
It estimates the components of the Weyl curvature by 
the quantity $\mathcal{Q}_1(W)$ introduced above. 
This quantity $\mathcal{Q}_1(W)$ is shown to be bounded. 
For the vectorfields $T$, $S$, $\bar{K}$, the corresponding deformation tensors are calculated. 
Here, the Bianchi equations play a crucial role. 
In fact, the Bianchi equations allow us to obtain the estimates of the angular derivatives of our curvature components 
directly. 
We re-emphasize that  no rotational vectorfields are needed in the present proof. 
In the work \cite{sta}, the authors introduced 
rotational vectorfields to obtain the corresponding angular derivatives. 
While this is different in our work, another fact is used similarly, that is the principle of conservation 
of signature. \\ \\ 
In the error estimates we show the  
quantity $Q_1(W)$, that is $Q_0(t)$ and $Q_1(t)$ to be bounded. 
In view of estimating $Q_1(W)$ from (\ref{intQ1W123}), we continue as follows. 
$Q_0(t)$ is controlled directly, once $Q_1(t)$ is estimated. 
The integral $Q_1(t)$ for $t_*$ can be split into 
\beas
Q_1 (t_*) \ & = & \ 
\int_{H_0} \ Q \ (\hat{\Lie}_S W) \ (\bar{K}, T , T, T) \\ 
& & \ + \ 
\int_{H_0} \ Q \ (\hat{\Lie}_T W) \ (\bar{K}, \bar{K} , T, T)  \\ 
& & \ + \ 
\mathcal{E}_1 (W, t_*) \ \ , 
\eeas
where 
$\mathcal{E}_1 (W, t_*)$ is the integral on $V_{t_*}$ of the absolute values of 
the error terms. $V_{t_*}$ denotes the spacetime slab $\bigcup_{t \in [0, t_*]} H_{t}$, 
for which the bootstrap assumptions hold. That is, the corresponding quantities 
are bounded by a small positive constant $\epsilon_0$. 
These error terms are generated because of the fact that the integral 
(\ref{intQ1W123}) on $H_t$ differs from an integral over $H_0$, as the 
vectorfields $T, S, K$ are not exact conformal Killing fields (only quasi-conformal). 
The expressions in $\mathcal{E}_1 (W, t_*)$ to be integrated are 
quadratic in the Weyl fields $W$ and linear in the deformation tensors $\hat{\pi}$ 
of the vectorfields. \\ \\ 
Generally, for the integral on $H_t$, we have the following formula, 
(see \cite{lydia2} and \cite{lydia1}, chapter 5, proposition 13), 
for three arbitrary vectorfields $X, Y, Z$ and $T$ denoting the unit normal to the foliation 
by a time function $t$: 
\beas
\int_{H_t} Q(W) (X, Y, Z, T) \ d \mu_g \ & = & \ 
\int_{H_0} Q(W) (X, Y, Z, T) \ d \mu_g  \\ 
& & 
+ \ 
\int_0^t \Big( 
\int_{H_{t'}} (div Q )_{\beta \gamma \delta} \ X^{\beta} \ Y^{\gamma} \ Z^{\delta}  \\ 
& & 
\ + \  
\frac{1}{2} \ \int_{H_{t'}} Q_{\alpha \beta \gamma \delta} \ 
\big( 
\ ^{(X)} \pi^{\alpha \beta} \ Y^{\gamma} \ Z^{\delta} \ + \ 
\ ^{(Y)}  \pi^{\alpha \beta} \ Z^{\gamma} \ X^{\delta} \\ 
& & 
\ \ + \  
\ ^{(Z)}  \pi^{\alpha \beta} \ X^{\gamma} \ Y^{\delta} \big) \ \Phi \  d \mu_g \Big) \ dt' \ \ . 
\eeas
We also need the following integral on the cones $C_u$ 
\bea
\tilde{Q}_1 (W, u, t) \ & = & \ \int_{C_u(t_0,t)} Q (\hat{\Lie}_S W) (\bar{K}, T, T, e_4) \nonumber \\
& & \ + \ 
\int_{C_u(t_0,t)} Q (\hat{\Lie}_T W) (\bar{K}, \bar{K}, T, e_4)   \label{intQ1WCu12}
\eea
with $C_u(t_0,t)$ denoting the part of the cone $C_u$ between 
$H_{t_0}$ and $H_t$ for $0 \leq t_0 \leq t$. \\ \\ 
And what we have just said for (\ref{intQ1W123}), also holds for (\ref{intQ1WCu12}). 
In fact, we estimate 
\be \label{intQ1*12}
Q_1^* \ = \ \max \ \Big( 
\sup_{t \in [0, t_*]} Q_1 (W, t) \ , \ \ \sup_{t \in [0, t_*]} \sup_{u_* \geq u_0(t)} 
\tilde{Q}_1 (W, u, t)  
\Big)  \ \ . 
\ee
Thus, we have the inequality 
\[
Q_1^* \ \leq \  Q_1(0) \ + \ \mathcal{E}_1 (t_*)  \ \ . 
\]
Therefore, we have to estimate the error terms 
$\mathcal{E}_1 (t_*)$. 
We prove by a bootstrap argument (see \cite{lydia2}, \cite{lydia1}, chapter 6, theorem 7) that 
\be \label{*estnew*1}
 \mathcal{E}_1 (t_*) \ \leq \ C  \epsilon_0  Q_1^* \ \ ,
\ee
which for $\epsilon_0$ sufficiently small implies: 
\be
Q_1^* \ \leq \ 2 \  Q_1(0) \ \ . 
\ee
In deriving (\ref{*estnew*1}) we encounter borderline estimates for the most delicate terms. \\ 
\subsection{Borderline Estimates}
We now give an example of a borderline estimate. To do so, let us first give the formula 
for the error terms. \\ \\ 
Let $V_t$ denote the spacetime slab $\bigcup_{t' \in [0, t]} H_{t'}$.  \\ \\ 
The first term in $Q_1(t)$ from formula (\ref{intQ1W123}) is 
\beas
& & 
\int_{H_t} \ Q \ (\hat{\Lie}_S W) \ (\bar{K}, T, T, T)   \  =  \ 
\int_{H_0} \ Q \ (\hat{\Lie}_S W) \ (\bar{K}, T, T, T)   \\ 
& & 
\ + \ 
\int_{V_{t}} 
\Phi \ (div \ Q(\hat{\Lie}_S W))_{\beta \gamma \delta} \ \bar{K}^{\beta} \ T^{\gamma} \ T^{\delta}  \\ 
& & 
\  + \ 
\frac{1}{2} \ \int_{V_{t}}  \Phi \ 
 Q(\hat{\Lie}_S W)_{\alpha \beta \gamma \delta} \  
 \ ^{(\bar{K})} \pi^{\alpha \beta} T^{\gamma} T^{\delta}  \\ 
& & 
\ + \ 
\int_{V_{t}}  \Phi \ 
 Q(\hat{\Lie}_S W)_{\alpha \beta \gamma \delta} 
 \ ^{(T)} \pi^{\alpha \beta} \bar{K}^{\gamma} T^{\delta}   \\ 
\eeas
Similarly, we obtain for the remaining terms: 
\beas
Q_1 (W, t) \ & \leq & \ Q_1 (W, 0) \ + \ \mathcal{E}_1 (W, t)  \\ 
Q_0 (W, t) \ & \leq & \ Q_0 (W, 0) \ + \ \mathcal{E}_0 (W, t) 
\eeas
with 
\beas
& & 
\mathcal{E}_1 (W, t)  \  =  \ 
\int_{V_t} \Phi \ \mid 
( div \ Q(\hat{\Lie}_S W))_{\beta \gamma \delta} \ \bar{K}^{\beta} \ T^{\gamma} \ T^{\delta} \mid  
\nonumber \\ 
& & 
+ \ \int_{V_t} \Phi \  \mid 
(div \ Q(\hat{\Lie}_T W))_{\beta \gamma \delta} \ \bar{K}^{\beta} \ \bar{K}^{\gamma} \ T^{\delta}  \mid  
\nonumber \\ 
& & 
+ \ 
\frac{1}{2} \ \int_{V_{t}} \Phi \  \mid  
 Q(\hat{\Lie}_S W)_{\alpha \beta \gamma \delta} \  
 \ ^{(\bar{K})} \pi^{\alpha \beta} T^{\gamma} T^{\delta} \mid  \nonumber \\ 
& & 
\ + \ 
\int_{V_{t}} \Phi \ \mid 
 Q(\hat{\Lie}_S W)_{\alpha \beta \gamma \delta} 
 \ ^{(T)} \pi^{\alpha \beta} \bar{K}^{\gamma} T^{\delta} \mid  \nonumber \\ 
& & 
\ + \ 
\int_{V_{t}} \Phi \ \mid 
 Q(\hat{\Lie}_T W)_{\alpha \beta \gamma \delta} \  
 \ ^{(\bar{K})} \pi^{\alpha \beta} \bar{K}^{\gamma} T^{\delta} \mid  \nonumber \\ 
& & 
\ + \ 
\frac{1}{2} \ \int_{V_{t}} \Phi \ \mid 
 Q(\hat{\Lie}_T W)_{\alpha \beta \gamma \delta} 
 \ ^{(T)} \pi^{\alpha \beta} \bar{K}^{\gamma} \bar{K}^{\delta} \mid \\ 
\eeas
\beas
& & 
\mathcal{E}_0 (W, t)  \  =  \ 
\int_{V_t} \Phi \ \mid \underbrace{(div \ Q(W))_{\beta \gamma \delta}}_{=0} \ \bar{K}^{\beta} \ T^{\gamma} \ T^{\delta} \mid  
\nonumber \\ 
& & 
+ \ 
\frac{1}{2} \ \int_{V_{t}}  \Phi \ \mid 
 Q(W)_{\alpha \beta \gamma \delta} \  
 \ ^{(\bar{K})} \pi^{\alpha \beta} T^{\gamma} T^{\delta} \mid  \nonumber \\ 
& & 
\ + \ 
\int_{V_{t}} \Phi \ \mid 
 Q(W)_{\alpha \beta \gamma \delta} 
 \ ^{(T)} \pi^{\alpha \beta} \bar{K}^{\gamma} T^{\delta} \mid
\eeas
There are several borderline cases appearing in 
$\mathcal{E}_1 (W, t_*)$. 
One of them we encounter in the integral  
\be \label{borderlineexbbb1}
\int_{V_{t_*}^e} \tau_+^2 \  \Phi \ \mid (\rho, \sigma) \ (\hat{\Lie}_S W) \mid \ 
\mid tr \chi \mid \ 
\mid \S \hat{i} \mid \ 
\mid  \underline{\alpha} \mid \ \ , 
\ee
which arises in the most delicate term 
\[
\int_{V_{t_*}^e} \Phi \tau_+^2 \ \mid 
( div \ Q(\hat{\Lie}_S W))_{334}  \mid 
\]
when estimating 
$\int_{V_{t_*}^e} \Phi \ \mid 
( div \ Q(\hat{\Lie}_S W))_{\beta \gamma \delta} \ \bar{K}^{\beta} \ T^{\gamma} \ T^{\delta} \mid $ from above. 
We stress the fact that $( div \ Q(\hat{\Lie}_S W))_{334}$ being multiplied by $\tau_+^2$ involves parts 
with the worst decay properties, which require more subtle estimates, as we shall see in the 
treatment of the borderline case (\ref{borderlineexbbb1}). 
Note that $V_{t_*}^e$ is the exterior region as introduced earlier. 
We can split the integrals in $\mathcal{E}_1 (W, t_*)$ and $\mathcal{E}_0 (W, t_*)$ into 
the interior region $V_{t_*}^i$ where $r \leq \frac{r_0}{2}$, and the exterior region $V_{t_*}^e$ where $r \geq \frac{r_0}{2}$. 
The interior integrals are estimated in a straightforward way, as in the interior all the components of the tensors 
$DW, \hat{\Lie}_T W, \hat{\Lie}_S W$ and all the components of the deformation tensors of the vectorfields 
$T, S, \bar{K}$ behave in the same manner. 
Whereas in the exterior, the components of the said tensors behave differently. Therefore, the 
exterior estimates are more complicated, as they depend on the structure of the nonlinear terms. \\ \\ 
Writing out the terms of $\int_{V_{t_*}^e} \Phi \tau_+^2 \ \mid 
( div \ Q(\hat{\Lie}_S W))_{334}  \mid$ in details, we find that the most delicate, namely the borderline terms are 
of the form (\ref{borderlineexbbb1}). 
The quantities we have to pay special attention to are 
$\underline{\alpha}$ and $\S i$. 
Here, $\underline{\alpha}$ is the null curvature component (\ref{intnullcurvalphaunderline*1}) 
and $\S i$ is the null component $\S \hat{\pi}_{AB}$ of the deformation tensor (see (\ref{intdeformationt*1})) 
tangent to the surface. 
In view of the coarea formula, we write 
the integral (\ref{borderlineexbbb1}) 
over $V_{t_*}^e$ as an integral on $C_u$ and an integral with respect to $u$. 
\[
\int_{V_{t_*}^e} \tau_+^2 \  \Phi \ \mid (\rho, \sigma) \ (\hat{\Lie}_S W) \mid \ 
\mid tr \chi \mid \ 
\mid \S \hat{i} \mid \ 
\mid  \underline{\alpha} \mid \ = \  
\int_{- \infty}^{u_*} du \int_{C_u} \tau_+^2 \  \Phi \ \mid (\rho, \sigma) \ (\hat{\Lie}_S W) \mid \ 
\mid tr \chi \mid \ 
\mid \S \hat{i} \mid \ 
\mid  \underline{\alpha} \mid
\]
We calculate on $C_u$: 
\beas
& & 
\int_{C_u} \tau_+^2 \  \Phi \ \mid (\rho, \sigma) \ (\hat{\Lie}_S W) \mid \ 
\mid tr \chi \mid \ 
\mid \S \hat{i} \mid \ 
\mid  \underline{\alpha} \mid \\ 
& &  \leq 
c' \ \sup_{C_u} \{ \tau_+ \mid tr \chi \mid \} \ 
\int_{C_u} \tau_+ \mid (\rho, \sigma) \ (\hat{\Lie}_S W) \mid \ 
\mid \S \hat{i} \mid \ 
\mid  \underline{\alpha} \mid \\ 
& &  \leq \ 
c 
\Big( \int_{C_u} \tau_+^2  \mid (\rho, \sigma) \ (\hat{\Lie}_S W) \mid^2 \ \Big)^{\frac{1}{2}} 
\Big( \int_{C_u} \mid \S \hat{i} \mid^2 \ 
\mid   \underline{\alpha} \mid^2 
\Big)^{\frac{1}{2}} 
\eeas
\[\]
Whereas the first integral on the right hand side is bounded by curvature assumptions, the second 
integral has to be investigated further. Thus, one obtains 
\bea
& & 
\Big( \int_{C_u} \mid \S \hat{i} \mid^2 \ 
\mid   \underline{\alpha} \mid^2 
\Big)^{\frac{1}{2}} 
\  =  \ 
\Big( 
\int_0^{t_*} 
\Big{\{}  
\int_{S_{t,u}} \mid \S \hat{i} \mid^2 \ 
\mid   \underline{\alpha} \mid^2 \Big{\}}   dt 
\Big)^{\frac{1}{2}}  \nonumber \\ 
& & 
\ = \ 
\Big( \ \int_0^{t_*} \ 
\parallel \ \ \mid \S \hat{i} \mid \ \mid \underline{\alpha} \mid \ \ \parallel_{L^2(S_{t,u})}^2 
\ dt \ \Big)^{\frac{1}{2}}   \nonumber \\ 
& & 
\  \leq  \ 
\Big( \ \int_0^{t_*} \ 
\parallel \S \hat{i} \parallel_{L^4(S_{t,u})}^2  \ 
\parallel \underline{\alpha} \parallel_{L^4(S_{t,u})}^2 
\ dt \ \Big)^{\frac{1}{2}}  \label{Bbdlcaserrest*1}
\eea
%
The problem reduces to estimating the last integral. 
We have 
\bea
& & 
\Big( \ \int_0^{t_*} \ 
\parallel \S \hat{i} \parallel_{L^4(S_{t,u})}^2  \ 
\parallel \underline{\alpha} \parallel_{L^4(S_{t,u})}^2 
\ dt \ \Big)^{\frac{1}{2}}   \nonumber  \\ 
& & 
\ \ \ \quad \leq  \ 
c \ \tau_-^{- \frac{3}{2}} \  \Big( \ \int_0^{t_*} \ 
(1 + t)^{-1} \ \parallel \S \hat{i} \parallel_{L^4(S_{t,u})}^2 
\ dt \ \Big)^{\frac{1}{2}} \ \ . \label{borderlineexbbb2}
\eea
The curvature assumptions give 
\[
\sup_{t, \ \mbox{\tiny on} C_u} \big{\{} 
r^{\frac{1}{2}} \ \parallel \underline{\alpha} \parallel_{L^4(S_{t,u})} \big{\}} \ \ \leq \ \ 
c \ \tau_-^{- \frac{3}{2}} 
\]
and the assumption on $ \S i$ is 
\[
\parallel r^{\frac{1}{2}} \S i \parallel_{\infty, e} \ \leq \ \epsilon_0 \ \ , 
\]
which gives 
\[
 \parallel \S \hat{i} \parallel_{L^4(S_{t,u})} \ \leq \ C \epsilon_0 \ \ . 
\]
If these bounds are substituted in (\ref{borderlineexbbb2}) a logarithmic divergence would result. 
Thus the integral in (\ref{borderlineexbbb2}) is borderline. 
We show that in fact it is bounded. 
Any relaxation of our assumptions on the data involved would make this integral diverge. 
Now, in view of the definition of $\S \hat{i}$ and writing out the 
Lie derivative 
\[
\hat{\Lie}_S g \ \ = \ \ 
\frac{1}{2} \ \underline{u} \ \hat{\Lie}_L g \ + \ 
\frac{1}{2} \ u \ \hat{\Lie}_{\underline{L}} g  \ \ , 
\]
the $\S \hat{i}$ involves the terms $r \hat{\chi}$ and $u \hat{\underline{\chi}}$. Whereas 
the latter has order of decay $O(r^{-1} \tau_-^{\frac{1}{2}})$, the first term is only of order $o(r^{- \frac{1}{2}})$. 
Therefore, by estimating $r \hat{\chi}$ we obtain bounds for $\S \hat{i}$. 
Using the Codazzi equations and 
the assumption that $\beta \in L^4(S_{t,u})$, 
we obtain by elliptic estimates from the results to control 
$r \hat{\chi}$ in terms of $\beta$: 
\bea
\parallel r \hat{\chi} \parallel_{L^4(S_{t,u})} \ + \ 
\parallel r^2 \nlap \hat{\chi} \parallel_{L^4(S_{t,u})} 
\ & \leq & \ 
c \ \parallel r^2 \beta \parallel_{L^4(S_{t,u})} \ \ .  \label{errestellest1}
\eea
As we want to integrate $(1 + t)^{-1} \parallel \S \hat{i} \parallel_{L^4(S_{t,u})}^2$ 
in (\ref{borderlineexbbb2}), we study the integral for \\ 
$\parallel r^{\frac{3}{2}} \beta \parallel_{L^4(S_{t,u})}^2$: 
\bea
\int_0^{t_*}  \parallel r^{\frac{3}{2}} \beta \parallel_{L^4(S_{t,u})}^2 \ dt 
 \ & \leq  & \  c \ 
\int_0^{t_*} \parallel r \beta \parallel_{L^2(S_{t,u})} \cdot \parallel r \beta \parallel_{L^2(S_{t,u})} \ dt  \nonumber 
 \\ 
& & 
 + \ c \ 
\int_0^{t_*} \parallel r \beta \parallel_{L^2(S_{t,u})} \cdot \parallel r^2 \nlap \beta 
\parallel_{L^2(S_{t,u})} \ dt  \label{errestproof***1} \\  
\ & \leq & \ c \ 
\Big( 
\int_0^{t_*} \parallel r \beta \parallel_{L^2(S_{t,u})}^2 dt \Big)^{\frac{1}{2}} \ 
\Big( 
\int_0^{t_*} \parallel r \beta \parallel_{L^2(S_{t,u})}^2 dt \Big)^{\frac{1}{2}}  \nonumber  \\ 
& & 
 + \ c \ 
\Big( 
\int_0^{t_*} \parallel r \beta \parallel_{L^2(S_{t,u})}^2 dt \Big)^{\frac{1}{2}} \ 
\Big( 
\int_0^{t_*} \parallel r^2 \nlap \beta \parallel_{L^2(S_{t,u})}^2 dt \Big)^{\frac{1}{2}}  \nonumber  \\ 
& &  \label{errestproof***4}  \\ 
\ & = & \
c \ 
\Big( 
\int_{C_u} r^2 \mid \beta \mid^2 \Big)^{\frac{1}{2}} \ 
\Big( 
\int_{C_u} r^2 \mid \beta \mid^2 \Big)^{\frac{1}{2}}   \nonumber  \\ 
& & 
 + \ c \ 
\Big( 
\int_{C_u} r^2 \mid \beta \mid^2 \Big)^{\frac{1}{2}} \ 
\Big(  
\int_{C_u} r^4 \mid \nlap \beta \mid^2 \Big)^{\frac{1}{2}}  \label{errestproof***2}
\eea
The right hand side of the last inequality is bounded by the curvature assumptions on $\beta$. 
In order to prove the first inequality, we apply the isoperimetric inequality on 
$S_{t,u}$ to $\mid \beta \mid^2$. 
Using the fact that for a function $f$ we have 
$\int_{S_{t,u}} (f - \bar{f})^2 = \int_{S_{t,u}} f^2 - \int_{S_{t,u}} f \bar{f} = \int_{S_{t,u}} f^2 - \int_{S_{t,u}} \bar{f}^2$ 
we obtain by first applying the isoperimetric and then H\"older inequality 
\beas
\int_{S_{t,u}} \mid \beta \mid^4 \ & \ \leq \ & \ 
c \ r^{-2} \  \Big( \int_{S_{t,u}}  \mid \beta \mid^2 d \mu_{\gamma}  \Big) \ 
\Big{\{}    
 \Big( \int_{S_{t,u}}  \mid \beta \mid^2 d \mu_{\gamma}  \Big) \ + \ 
\Big(  \int_{S_{t,u}}  r^{2} \mid \nlap \beta \mid^2 d \mu_{\gamma} \Big)  
\Big{\}}
\eeas
Multiplying by $r^6$ we derive 
\be \label{errestproof***3}
\parallel r^{\frac{3}{2}} \beta \parallel_{L^4(S_{t,u})}^2 \ \leq \ 
C \ \Big{\{}  
\parallel r \beta \parallel_{L^2(S_{t,u})}^2 \ + \ \parallel r \beta \parallel_{L^2(S_{t,u})} \parallel r^2 \nlap \beta  \parallel_{L^2(S_{t,u})} 
\Big{\}}
\ee
Finally, 
estimate (\ref{errestproof***1}) for the integral over $t$ is deduced, and 
the Cauchy-Schwarz inequality then gives the right hand side of (\ref{errestproof***4}) respectively (\ref{errestproof***2})  
which is bounded. 
What we have shown is 
\beas
\parallel r^{- \frac{1}{2}} \S \hat{i} \parallel_{L^2([0,t_*], L^4(S_{t,u}))}  
\ & \leq & \ 
c \ 
\Big( 
\int_{C_u} r^2 \mid \beta \mid^2 \Big)^{\frac{1}{2}} \ 
\Big( 
\int_{C_u} r^2 \mid \beta \mid^2 \Big)^{\frac{1}{2}}  \\ 
& & 
 + \ c \ 
\Big( 
\int_{C_u} r^2 \mid \beta \mid^2 \Big)^{\frac{1}{2}} \ 
\Big( 
\int_{C_u} r^4 \mid \nlap \beta \mid^2 \Big)^{\frac{1}{2}}  \\ 
\ & \leq & \ 
c \ \epsilon_0 \ Q_1^*  \ \ . 
\eeas
From this it directly follows 
\[
\int_{V_{t_*}^e} \tau_+^2 \ \Phi \ \mid  (\rho, \sigma) \ (\hat{\Lie}_S W) \mid \ 
\mid tr \chi \mid \ 
\mid \S \hat{i} \mid \ 
\mid  \underline{\alpha} \mid 
\ \ \leq \ \ c \ \epsilon_0 \ Q_1^* \ \ . 
\]
We observe that only the estimates of $\S i$ in terms of $\beta$ yield  
the required bounds for this integral. Any further relaxation of our assumptions would lead 
to divergence of this integral. Therefore, this is indeed a borderline case. \\ \\ 
The estimates for many quantities in our work are borderline. Let us point out here, that this is true also for 
$\chi$. Most of these borderline cases 
require a more careful treatment. \\ 
\subsection{Controlling Geometric Quantities and the Construction of the Optical Function}
The main step 3 above, consists of  
estimating 
the geometric quantities with respect to the foliations $\{ H_t \}$ and $\{ C_u \}$. 
The core part in here is the treatment of 
the components of the second fundamental form $k$ 
and the Hessian of $u$. 
In view of the results mentioned in the previous paragraph, 
the estimates derived here, have to be appropriate. Then, this makes it possible to close the proof 
of the main theorem and to derive the main results. 
To make this more precise, let us discuss in the following the procedure how it is done, and 
give the different quantities that are estimated. \\ \\ 
We introduce the basic norms of our geometric quantities in subsection \ref{intnorms*a1}, that is in particular, 
of the components of $k$ in \ref{norms*secondfundffk**subsection*1} 
and of the components of the Hessian of $u$ in \ref{norms*hessoptfctu**subsection*1}.  
With the definitive version of our main theorem \ref{maintheoremlb*1} 
comes a concrete description of the 
asymptotic behaviour of the geometric quantities. \\ \\ 
The second fundamental form $k$ 
decomposes into the scalar $k_{NN} = \delta$, 
the $S$-tangent $1$-form $k_{AN} = \epsilon_A$ and 
the $S$-tangent, symmetric $2$-tensor $k_{AB} = \eta_{AB}$. 
The worst decay properties have $\hat{\eta}_{AB}$, that is, the traceless part of $\eta_{AB}$. 
Having good estimates for the curvature, we can then use standard tools. 
Here, elliptic estimates are applied to derive the desired results. 
The main part is to prove them in the wave zone, where  
the behaviour of the components and of their derivatives depends on the direction. 
The elliptic system on $H_t$ for $k$ is given by: 
\bea
tr k \ & = & \ 0 \label{ktrk0} \\
curl \ k \ & = & \ H \label{kcurlkH} \\
div \ k \ & = & \ 0 \label{kdivk0} \ . 
\eea
$H$ is the magnetic part of the spacetime curvature relative to the time foliation. 
We also have: 
\be
\bar{R}_{ij} \ = \ k_{ia} \ k^a_{\ j} \ + \ E_{ij} \ , 
\ee
where $E$ denotes the electric part of the spacetime curvature relative to the time foliation. 
This elliptic system on $H_t$ for $k$ is decomposed relative to the radial foliation, that is the foliation of 
each $H_t$ by the surfaces $S_{t,u}$. 
From the fact, that the nonlinear terms in these equations behave better in view of decay than the 
worst linear ones, follows that the estimates are essentially linear but 
nevertheless yield control of the full nonlinear problem. 
The estimates, being essentially linear in the present situation in contrast to 
\cite{sta}, simplify the proof considerably. \\ \\ 
In order to study the components of the Hessian of $u$, 
we apply the basic method from the original proof in \cite{sta}, namely 
the method of treating propagation equations along 
the cones $C_u$ coupled to elliptic systems on the surfaces $S_{t,u}$. 
However, our estimates differ fundamentally from the ones in \cite{sta}. 
The reason for that is the fact that we do not have any $L^{\infty}$ bounds on the curvature, but 
we only control one derivative of the curvature 
in the hypersurfaces $H_t$ and the Gauss curvature $K$ in the surfaces $S_{t,u}$ lies in 
$L^4$, as explained above. Our situation yields borderline estimates for $\chi$ (see below), whereas in 
\cite{sta}, there are no borderline estimates. In fact, our assumptions on the fall-off cannot be 
relaxed. Thus, they are sharp with respect to decay. \\ \\ 
Relative to a null frame (introduced before) the Hessian of $u$ decomposes into 
$\chi_{AB}$, $\zeta_A$, $\omega$, satisfying the following equations: 
\beas
\frac{d \chi_{AB}}{ds} \ & = & \ 
- \ \chi_{AC} \ \chi_{CB} \ + \ \alpha_{AB} \\ 
\frac{d \zeta_A}{ds} \ & = & \ 
- \ \chi_{AC} \ \zeta_B \ + \ \chi_{AB} \ \underline{\zeta}_B \ - \ \beta_A \\ 
\frac{d \omega}{ds} \ & = & \ 
2 \ \underline{\zeta} \cdot \zeta  \ - \ \mid \zeta \mid^2 \ - \ \rho \ . 
\eeas
As $\alpha$ is traceless, the trace of $\chi$, fulfills an equation without curvature terms, 
namely (\ref{dtrchi}): 
\[
\frac{d tr \chi}{ds} \ = \ - \ \frac{1}{2} \ (tr \chi)^2 \ - \ \mid \hat{\chi} \mid^2  
\]
with $\hat{\chi}$ denoting the traceless part of $\chi$. 
For $\hat{\chi}$ the null-Codazzi equations form an elliptic system on the surfaces $S_{t,u}$. Recall 
(\ref{chihatfa}), which in details read: 
\be
\dlap \ \hat{\chi}_a 
\  =  \ 
-  \beta_a 
 \ + \ \frac{1}{2} \nlap_a   tr   \chi \ - \ \hat{\chi}_a^{\ b} \ \zeta_b \ + \ \frac{1}{2} \ 
tr  \chi \ \zeta_a  \ .  \label{intdchihat}
\ee
$\nlap$ is the induced covariant differentiation on the surfaces $S_{t,u}$. 
As a main goal here, we show that $\chi$ is one degree of differentiability smoother than 
the curvature. This is achieved by a bootstrap argument with a certain assumption on the curvature 
term $\beta$ on the right hand side of (\ref{intdchihat}), 
as sketched above. The divergence equation (\ref{intdchihat}) 
is a Hodge system. 
Thus, the bootstrap argument together with (\ref{dtrchi}), (\ref{intdchihat}), the method of 
treating such coupled systems of propagation and elliptic equations together with  
the Hodge theory yield the estimates. 
In the same line, we obtain the results for $\zeta$ and $\omega$. For 
$\zeta$, we have the Hodge system 
\beas
\dlap  \zeta \ & = & \ 
- \ \mu \ - \ \rho \ + \ \frac{1}{2} \ \hat{\chi} \ \cdot \ \underline{\hat{\chi}}  \\ 
\clap \ \ \zeta \ & = & \ 
\sigma \ - \ 
\frac{1}{2} \ \hat{\chi} \ \wedge \ \underline{\hat{\chi}}   
\eeas
with $\underline{\chi}$ being the second fundamental form 
\[
\underline{\chi} (X, Y) \ =  \ g (D_X \underline{L}, Y) 
\]
for $X, Y \in T_pS$ and $\underline{L}$ being the inward null normal 
whereas $\mu$ denoting the 
\begin{itshape}mass aspect function \end{itshape}
\[
\mu \ = \ K \ + \ \frac{1}{4} \ tr \chi \ tr \underline{\chi} \ - \ 
\dlap \zeta \ . 
\]
The propagation equation for 
$\mu$ is derived to be 
\bea
\frac{d \mu}{ds} \ + \ \frac{3}{2} \ tr \chi \ \mu  \ & = & \ 
\ - \ \frac{1}{4} \ tr \underline{\chi} \ \mid \hat{\chi} \mid^2 
\ + \ 
\frac{1}{2} \ tr \chi \ \mid \zeta \mid^2  \nonumber  \\ 
& & \ + \ 
2 \ \dlap \chi \cdot \zeta 
\ + \ 
\hat{\chi} \cdot \nlap \hat{\otimes} \zeta  \ \ , 
\eea
with 
$(\nlap \hat{\otimes} \zeta)_{ab} = \nlap_a \zeta_b + \nlap_b \zeta_a - \gamma_{ab} \ \dlap \zeta$. \\ \\ 
Wihtin this framework, to obtain our estimates, the uniformization theorem for 
$K \in L^4(S)$ is needed, as it is mentioned above and proven in \cite{lydia1}, \cite{lydia2}.  \\ \\ 
A further crucial part of our main proof deals with the 
last slice $H_{t_*}$ of the spacetime. 
Here, we study the construction of the optical function $u$ on this last slice. 
It is done by solving an inverse lapse problem for the function $u$ on $H_{t_*}$. 
That is, starting from 
$S_{t_*, 0} = H_{t_*} \cap C_0$, 
the intersection of the last slice with the initial cone $C_0$, the function 
$u$ is defined to be the solution of: 
\[
\mid \nabla u \mid^{-1} \ = \ a \ \ , \ \ \ \ \quad \ u \ \mid_{S_{t_*,0}} \ = \ 0  \ , 
\]
where $a$ on each $S_{t_*,u}$ fulfills 
\[
\slap \log a \ = \ f \ - \ \bar{f} \ - \ \dlap \epsilon \ \ , \ \ \ \ \quad \ \overline{\log a} \ = \ 0 
\]
with 
\[
f \ = \ K \ - \ \frac{1}{4} \ (tr \chi)^2 \  . 
\]
This method of solving an inverse lapse problem for $u$ is the same as in 
\cite{sta}, whereas the proof itself differs fundamentally from the one in \cite{sta}. 
Again the fact that we do not have any $L^{\infty}$ bounds on the curvature is crucial, but 
the curvature only 
lies in $H^1$ of the hypersurface $H_{t_*}$. \\ \\ 
Constructing $u$ as a solution of the inverse lapse problem, that is solving an equation of motion of surfaces, 
was first done by D. Christodoulou and S. Klainerman in \cite{sta}. 
One might first think of applying other, easier methods in order to construct $u$. 
But, as we are going to explain now, they would not match our requirements. \\ \\ 
What we refer to as the `last slice' is the maximal hypersurface  $\mathcal{H}_{t_*}$, which 
bounds in the future the spacetime slab that we are constructing in the continuity argument. 
The obvious choice of $u$ on  $\mathcal{H}_{t_*}$, namely minus the 
signed distance function from $S_{t_*,0}$, is inappropriate because 
this distance function is only as smooth as the induced metric 
$\bar{g}_{t_*}$. It is not one order better, which would be the maximal possible. 
Thus, there would be a loss of one order of differentiability. But the 
problem does not allow to lose derivatives. That is, 
with this loss of one order of differentiability, the estimates 
would fail to close.  \\ \\ 
To overcome this difficulty, we define $u$ on $\mathcal{H}_{t_*}$ in a 
different way, namely by solving an equation of motion of surfaces on 
$\mathcal{H}_{t_*}$, as described above. \\ \\ 
Why would the use of other methods like the inverse mean curvature flow (IMCF) not work here? 
Using IMCF, the problem could be solved in the outward direction only. Whereas the 
equation of motion of surfaces can be solved in both directions, which is what we need. \\ \\ 
The equation, we use here, has to have the smoothing property described above as well as it has to be 
solvable in both directions. This excludes the IMCF and similar methods. 
It turns out that the equation of motion of surfaces yields exactly what we need. \\ \\ 
Let us explain now the principal ideas in the proof of the main theorem in the last slice and 
also show, in which sense our relaxed assumptions on the curvature require a special treatment.  \\ \\ 
The main results 
of this part are again obtained by a bootstrap argument. 
We have to assume estimates for the spacetime curvature on the last slice $H_{t_*}$. 
To be precise, these assumptions have to be made with respect to the background foliation, as 
the level surfaces of $u$ on the last slice do not yet exist, but have to be constructed. Note that, 
at the beginning, only $S_{t_*, 0}$ is given. 
We have to use the estimates on the background foliation to 
control the curvature and geometric components of the foliation by $u$. 
We estimate $tr \chi$, $\hat{\chi}$, $\zeta$ and their first angular derivatives, as well as 
$K$ in dimensionless $L^4$-norms on the surfaces $S_{t_*, u}$. We show this by bootstrapping. 
A crucial role in the proof is played by the trace lemma, yielding $L^4$-bounds on the 
curvature components in $S_{t_*, u}$. Relying on them, we then apply elliptic theory 
to obtain the estimates of the theorem. By a straightforward argument, the said quantities  
are shown to be controlled correspondingly in $H_{t_*}$. 
Next, integrating in the last slice over $u \in [0, \infty)$, yields 
the estimates for the second angular derivatives of $tr \chi$, $\hat{\chi}$ and $\zeta$ in 
$L^2$-norms in $H_{t_*}$. \\ \\ 
Yet, in order to apply the trace lemma, as said above, one has to work more. 
We estimate on $H_{t_*}$ the components of the above quantities of the foliation by the function $u$ by 
corresponding quantities with respect to the background foliation given by $u'$. 
Let $u'$ be a smooth function without critical points, defined in a 
tubular neighbourhood $U'$ of $S_0$, and let 
$S'_{u'}$ be the level sets of $u'$ on $H_{t_*}$ of our background foliation. 
Denote 
\[
U' \ = \ \bigcup_{u' \in (u'_0, \infty)} S'_{u'} \ \ . 
\]
The function $u'$ is introduced and specified in details 
in \cite{lydia2} and in \cite{lydia1}
in chapter 11, theorem 6. 
The curvature components with respect to its foliation are small. 
However, it is not clear if this function fulfills the equations of the inverse lapse problem above. 
Therefore, we use the estimates on the background foliation to control the curvature and geometric components of the foliation by $u$. 
For the foliation in the bootstrap argument, with respect to the function $u$, 
we denote 
\[
U \ = \ \bigcup_{u \in [0, u_1)} S_u \ \ . 
\]
We have to overcome the following difficulty: 
By assumption, the curvature components of the background foliation lie in 
$H^1(U')$. How can they be bounded in $H^1(U)$? 
There is no straightforward procedure to bound the curvature components 
in the surfaces of the foliation given by $u$ directly, as it is done in \cite{sta}, where 
the corresponding curvature components of the background foliation are in $L^{\infty}$. 
Thus the situation here is different. 
We solve the problem as follows: 
The transformation coefficients between quantities referring to the two foliations being bounded, 
we derive that 
the curvature components relative to the $u$-foliation lie in $H^1 (U)$. 
Only now, the trace lemma can be applied to obtain the curvature components with respect to the 
$u$-foliation to lie in $L^4 (S_u)$. We re-emphasize, that,  
as a consequence from controlling one less derivative, 
the derivatives of these components 
are not bounded in the surfaces $S_u$.  \\ \\ 
The main results of the last slice are formulated in $L^4$-norms. 
Actually, they hold for $L^p$-norms with $2 < p \leq 4$. The upper bound $4$ is given 
by the trace lemma, the lower bound $2$ by the fact that at certain levels of the proof, in the surfaces 
$S_{t_*,u'}$, we have to bound the $L^{\infty}$-norms of the quantities we estimate, and we only have 
them in $L^p$ up to their first derivatives, we have to require $p > 2$. 
This is necessary in view of the fact that for the said surfaces it is 
$W^p_m \hookrightarrow L^{\infty}$ for $mp >2$. \\ \\ 
Finally, all the bootstrap arguments close and so does the overall bootstrap argument, 
which finishes the proof of the main theorem. \\ \\ \\ 
Conlcuding, we find that the estimates for some of our main quantities are borderline from the point of view 
of decay, which means that it is not possible to relax further our assumptions. 
This indicates that the conditions in our theorem are sharp in so far 
as the assumptions on the decay at infinity on the initial data are concerned.  \\ \\ \\ 
{\bf Acknowledgment.} The results of my work \cite{lydia1}, \cite{lydia2}, on which this paper is based are 
from my Ph.D. thesis at the Swiss Federal Institute of Technology (ETH) Zurich. 
I would like to express my deepest gratitude to my thesis advisor Demetrios Christodoulou 
for his constant guidance and support and for many interesting discussions. 
Especially I would like to thank him for his ideas that enter my thesis and monograph, on estimating 
$\chi$, uniformization and for the idea of not using rotational vectorfields at all. 
I also thank Michael Struwe for helpful comments on my original work and for having been the co-advisor of my thesis. 
I thank especially S.-T. Yau for interesting discussions and his encouragement to publish 
these results.

\newpage 


\end{document}